\documentclass[twocolumn]{aa}
\usepackage[varg]{txfonts}

\usepackage{graphicx}
\usepackage{textcomp}
\usepackage{rotating}
\usepackage{color,xcolor}
\usepackage{bbold}
\usepackage{breqn}
\usepackage{amsmath}
\usepackage{pstricks} 
\usepackage{bm} 
\usepackage{relsize}
\usepackage{footmisc}
\usepackage{breqn}
\usepackage[colorlinks=true,urlcolor=blue,citecolor=blue,linkcolor=blue,breaklinks=true]{hyperref}
\usepackage{twoopt}
\DeclareMathOperator{\sign}{sign}

\def\xlinkspace#1 #2{%
 \ifx\relax#2%
 \xlinkdash#1-\relax
 \else
 \xlinkdash#1 -\relax
 \expandafter\xlinkspace\expandafter#2%
 \fi}

\def\xlinkdash#1-#2{%
 \ifx\relax#2%
 \tmp{#1}%
 \else
 \tmp{#1-}%
 \expandafter\xlinkdash\expandafter#2%
 \fi}

\bibpunct{(}{)}{;}{a}{}{,} % to follow the A&A style

\makeatletter

 \newcommandtwoopt{\citeads}[3][][]{%
   \nonstopmode%              %% fix to not stop at error message in latex
   \href{http://adsabs.harvard.edu/abs/#3}%
        {\def\hyper@linkstart##1##2{}%
         \let\hyper@linkend\@empty\citealp[#1][#2]{#3}}%   %% Rutten, 2000
   \biblink{#3}{\href{http://adsabs.harvard.edu/abs/#3}{ADS}}%
   \errorstopmode}            %% fix to resume stopping at error messages
   
 \newcommandtwoopt{\citepads}[3][][]{%
   \nonstopmode%              %% fix to not stop at error message in latex
   \href{http://adsabs.harvard.edu/abs/#3}%
        {\def\hyper@linkstart##1##2{}%
         \let\hyper@linkend\@empty\citep[#1][#2]{#3}}%     %% (Rutten 2000)
   \biblink{#3}{\href{http://adsabs.harvard.edu/abs/#3}{ADS}}%t
   \errorstopmode}            %% fix to resume stopping at error messages
   
 \newcommandtwoopt{\citetads}[3][][]{%
   \nonstopmode%              %% fix to not stop at error message in latex
   \href{http://adsabs.harvard.edu/abs/#3}%Section
        {\def\hyper@linkstart##1##2{}%
         \let\hyper@linkend\@empty\citet[#1][#2]{#3}}%     %% Rutten (2000)
   \biblink{#3}{\href{http://adsabs.harvard.edu/abs/#3}{ADS}}%
   \errorstopmode}            %% fix to resume stopping at error messages
   
 \newcommandtwoopt{\citeyearads}[3][][]{%
   \nonstopmode%              %% fix to not stop at error message in latex
   \href{http://adsabs.harvard.edu/abs/#3}%
        {\def\hyper@linkstart##1##2{}%
         \let\hyper@linkend\@empty\citeyear[#1][#2]{#3}}%  %% 2000
   \biblink{#3}{\href{http://adsabs.harvard.edu/abs/#3}{ADS}}%
   \errorstopmode}            %% fix to resume stopping at error messages
\makeatother

\makeatletter
\newcommand{\bibnote}[2]{\@namedef{#1note}{#2}}
\newcommand{\biblink}[2]{\@namedef{#1link}{#2}}
\makeatother

\newcommand{\be}{\begin{equation}}
\newcommand{\ee}{\end{equation}}

\newcommand{\bea}{\begin{eqnarray}}
\newcommand{\eea}{\end{eqnarray}}

\begin{document}
%\title{Augmentation of the S/N ratio of full-Stokes polarimetry through spatio-temporal regularisation}
\title{Characterization of telecentric dual-etalon Fabry-P\'erot systems from observational data}
\subtitle{Properties of the CRISP2 instrument at the Swedish 1-m Solar Telescope}
\titlerunning{An FPI characterization method}
\authorrunning{de la Cruz Rodr\'iguez et al.} 

%\subtitle{Bla}

\author{
  J. de la Cruz Rodr\'{i}guez \and
  G. B. Scharmer \and
  P. S\"utterlin \and
  J. Leenaarts \and
  M. G. L\"ofdahl\and
  D. Kiselman\and
  T. Hillberg\and
  O. Andriienko
}

\offprints{J. de la Cruz Rodr\'iguez \email{jaime@astro.su.se}}

\institute{Institute for Solar Physics, Dept. of Astronomy, Stockholm University, AlbaNova University Centre, SE-106 91 Stockholm, Sweden}

\date{Received; Accepted }

\abstract 
%context
{Imaging Fabry-P\'erot Interferometer (FPI) observations are commonly used in solar physics to infer physical parameters in the photosphere and chromosphere through modeling of the observations. Such techniques require detailed knowledge of the spectral instrumental profile in order to produce accurate results.}
{In this study we present a method to characterize the spatial variation of parameters of dual-etalon FPI instruments mounted in telecentric configuration: spatially-resolved cavity separation and reflectivities of both etalons, as well as the prefilter variation across the field-of-view. Here, we aim at characterizing the field-of-view dependence of the parameters of the new CRISP2 FPI.}
{We have implemented a forward model of the FPI instrumental degradation combined with a template average quiet-Sun  {spectrum} at disk center in order to model two sets of observational data.  Our method does not require any change in the optical setup or the utilization of external sources of illumination. We assess the validity of several functional forms in the calculation of the FPI transmission profiles.}
{Our results show that (generally) the inclusion of the secondary transmission peaks at $\pm 1$ times the Free Spectral Range and a detailed estimate of the prefilter curve is necessary in order to obtain accurate values of both etalon reflectivities. For  {very} narrow prefilters (relative to the FSR), the former requirement can be relaxed. Our results show that the cavity separation of CRISP2 is very flat, with an RMS variation below $2$~nm over the entire field-of-view for both etalons. Reflectivity RMS variations are $0.4\%$ and $0.3\%$ for the primary and secondary etalons at $617.3$~nm.}
{We have assessed data and modeling requirements to derive accurate  {dual-etalon telecentric} FPI parameters and minimize errors in the determination of etalon reflectivities.  {The} methods  {described in this paper} are relevant for the characterization of present and future FPI instruments  {and we have made them publicly available to the solar community}.  }
\keywords{Instrumentation: interferometers; Techniques: imaging spectroscopy; Methods: data analysis}

    \maketitle

\nolinenumbers

\section{Introduction} \label{sec:intro}
 {The use of FPIs in imaging polarimeters have become increasingly popular in observational solar physics during} the past two decades. Allowing for very narrow band imaging ($R = \lambda/\delta\lambda \approx 50000$--$120000$), these instruments can reconstruct a spectral line profile in each pixel of the field-of-view (FOV) by sequentially scanning through the wavelengths of the profile. Their great advantages are that they allow for  {a large FOV} (e.g., \citeads{2026A&A...705A..55S}), that image reconstruction techniques can be easily applied to FPI observations (\citeads{2005SoPh..228..191V}), and that a polarizing beam splitter, needed for polarimetric measurements, can be mounted close to the science cameras. 

Arguably, their main limitation is related to the limited amount of time that can be spent in the spectral scan before the solar scene cannot be assumed to be unchanged between the different wavelength positions of the scan.  {The latter is largely influenced by the spatial resolution of the system, as well as the characteristic evolution time of the layer of the solar atmosphere under study. For example, the scanning time for chromospheric observations acquired with a 1-m solar telescope (with a critical sampling of $47$~km~px$^{-1}$ at 617.3~nm) can be dominated by the Alfv\'en speed ($\sim 100$~km~s$^{-1}$) or by the sound speed ($\sim 10$~km~s$^{-1}$)}, which essentially translates into having a fixed amount of time ($\la\!\!  1 -10$ s) to acquire a line scan (\citeads{2006ApJ...648L..67V}; \citeads{2018A&A...614A..73F}; \citeads{2023A&A...669A..78S}; \citeads{2023A&A...673A..11R}).  {At lower spatial resolution this requirement is less constrained by fast dynamics in the solar atmosphere and longer scan times can be adopted}. The scan time can be spent in  {a} few line positions with long exposure time or many  positions with shorter exposure time. Examples of modern FPIs are IBIS (\citeads{2006SoPh..236..415C}), CRISP/CRISP2/ {CHROMIS} (\citeads{2006A&A...447.1111S}; \citeads{2026A&A...705A..55S}), VTF (\citeads{2017JATIS...3d5002S}), PHI (\citeads{2020A&A...642A..11S}),  {and TuMag (\citeads{2025arXiv250208268D})}.

FPIs have posed one of the best alternatives for photospheric and chromospheric studies in which a large FOV and relatively high cadence could be required. But in order to model FPI datasets, usually utilizing data \emph{inversion} techniques (\citeads{2016LRSP...13....4D}), an accurate estimate of the  {spectral} instrumental  {transmission} profile  {(transmission profile hereafter)} is required to make a meaningful comparison  {of synthetic spectra from simulations or inversion methods} with the observations.  {Unavoidable instrumental} effects for etalons mounted in telecentric configuration are field-dependent (random) wavelength shifts and variations in the FWHM of the transmission profile. These effects can be calibrated in order to produce a FOV-dependent spectral  {transmission profile}, that should be included in the inversion process.

 {A recent series of papers (\citeads{2019ApJS..241....9B}; \citeads{2019ApJS..242...21B};  \citeads{2020ApJS..246...17B}; \citeads{2021ApJS..254...18B}) have  {derived analytical expressions for the characterization of single etalon systems. Such expressions allow for a very fast determination of the instrumental profile of an etalon. However, the validity of analytical expressions  is restricted to cases where the etalon is not tilted, which is a crucial aspect of dual etalon systems for removing inter-etalon reflections.}}

We describe a FPI characterization method for telecentric dual-etalon systems (consisting of one prefilter and two etalons) using observational datasets. The techniques described in this paper build upon work presented in previous studies (\citeads{2006A&A...447.1111S}; \citeads{delacruz2010}; \citeads{2019ApJS..241....9B}; \citeads{2024A&A...688A..67S}) and over many years of experience at the SST. Similar techniques have also been used in the characterization of FPI instruments of collimated configurations (\citeads{2008A&A...481..897R}). Compared to previous studies,  {we include a model that can deal with the symmetry-breaking effects of a tilted etalon, a full characterization of the order-suppressing etalon, and} an assessment of the importance of accurate prefilter modeling.

The methods are utilized to analyze the performance of the new CRISP2 FPI (\citeads{2026A&A...705A..55S})
mounted at the Swedish 1-m Solar Telescope (SST, \citeads{2003SPIE.4853..341S}) at 617.3~nm. CRISP2 is an evolution of CRISP (\citeads{2006A&A...447.1111S}), optimized for a larger FOV.

\section{The CRISP2 FPI system}

\begin{figure}
	\centering
	\includegraphics[width=\columnwidth]{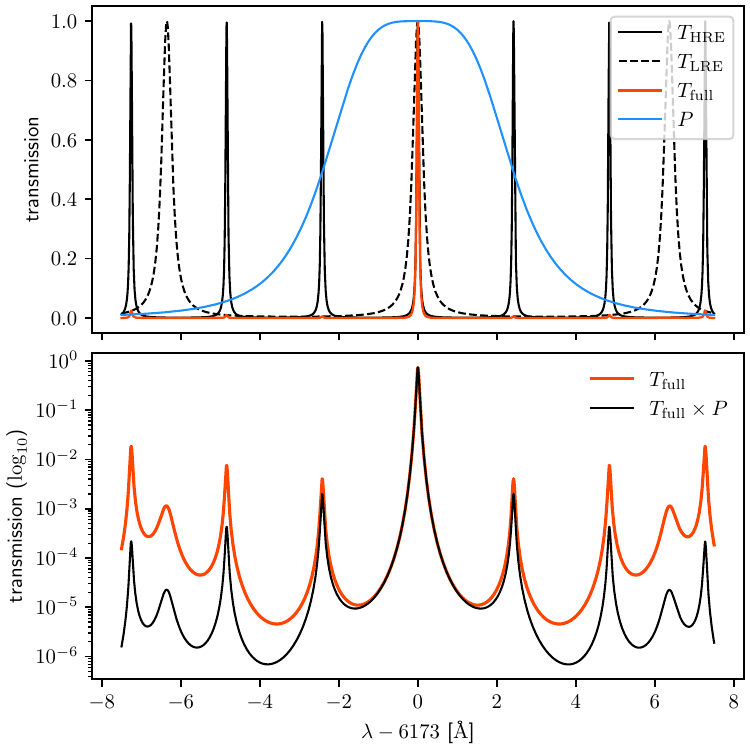}
	\caption{CRISP2 transmission profile at 617.3~nm. \emph{Top}: HRE, LRE and full transmission profiles, where the prefilter shape is also indicated. \emph{Bottom}: transmission profile and transmission profile multiplied by the prefilter transmission, illustrating the attenuation of secondary lobes by the prefilter. }\label{fig:tr_crisp2}
\end{figure}

The design  {concept} of CRISP2 \citetads{2026A&A...705A..55S} is identical to that of CRISP in that it consists of two etalons mounted in telecentric configuration  {and with the same} cavity separation ratio  {as CRISP. However, CRISP2 is} optimized for a larger FOV ($D_{\mathrm{fov}}\approx 120\arcsec$) than CRISP by means of  {larger etalons and} a lower F-ratio (F/140)  {at the etalons}. A high-spectral resolution etalon (HRE hereafter), having a narrow passband, sets the wavelength of the observation. A second etalon with a carefully chosen
(and different) free-spectral range (the spectral distance between consecutive transmission peaks), effectively suppresses secondary transmission peaks, leaving only the main transmission peak at the desired wavelength (see Fig.~\ref{fig:tr_crisp2}).  {To mitigate the negative effects of the lower F-ratio, the high-resolution etalon of CRISP2 is designed to have a slightly lower reflectivity (93\%) than that of CRISP (94\%).}

\citetads{2006A&A...447.1111S} proposed to use a low resolution secondary etalon (LRE) with a lower reflectivity than that of the primary etalon. The lower reflectivity is used  { to widen its transmission profile, which} strongly reduces the effect of  {mismatch of} the profiles of the two etalons {, originating from} random errors in their cavity separations. This simple trick helps keeping the system transmission high, while also minimizing  {variations in the shape} of the transmission profile across the FOV. An order-selecting prefilter is placed before the etalons to suppress light outside the spectral window under consideration and the contribution from secondary transmission peaks.  {Internal reflections between the two etalons are eliminated by tilting the LRE and placing a pupil stop at the location of the pupil image in front of the camera lens. Therefore, phase errors terms from inter-etalon reflections are completely eliminated in this type of design.}

\begin{figure}
	\centering
	\includegraphics[width=\columnwidth]{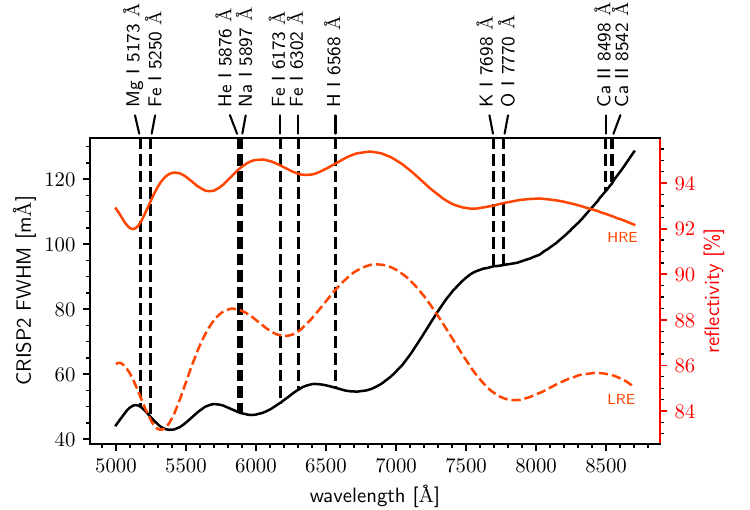}
	\caption{Factory-measured reflectivities of the etalons (red) and the corresponding instrumental profile FWHM for the CRISP2 instrument (black). The solid red line depicts the reflectivity of the high-resolution etalon, whereas the dashed-red line corresponds to the reflectivity of the low-resolution etalon. The FWHM of the profile was calculated assuming perfect co-tuning of the profiles from the two etalons and the factory nominal reflectivities.}\label{fig:ref}
\end{figure}

The nominal cavity separation of the two etalons are $C_{\mathrm{HRE}}^0=787$~$\mu$m and $C_{\mathrm{LRE}}^0=300$~$\mu$m, resulting in a cavity separation ratio (co-tuning factor) of $r_c = C_{\mathrm{LRE}}^0/C_{\mathrm{HRE}}^0 = 0.38119$. Our measured co-tuning factor is very similar $r_c= 0.38273$. Fig.~\ref{fig:ref} illustrates the factory-measured reflectivities of the two etalons and the resulting full-width-half-maximum (FWHM) of the transmission profile as a function of wavelength. Table~\ref{tab:ref} contains a list of spectral lines that have been or could be observed with CRISP2 and the corresponding reflectivities and transmission profile FWHM.

%The etalons have a clear aperture of $100$~mm, allowing for a maximum circular FOV of $120$\arcsec. 
%The prefilter diameter is $50$~mm. At the time of writing, we only had two of these larger prefilters, $656.3$~nm, and $854.2$~nm. A %$617.3$~nm prefilter was being manufactured during the preparation of this manuscript. The smaller CRISP prefilters  ($d=35$~mm) can be %utilized with a minimal FOV reduction. We have utilized a 35~mm prefilter, so the very outer ring of the FOV has not been covered. 

To make use of the full potential FOV of CRISP2, 50-mm-diameter filters (clear aperture > 40 mm) are used. At the time of writing, we only had two of these larger prefilters, designed for H-$\alpha$ 656.3~nm, and Ca\,{\sc ii} 854.25~nm with a filter for Fe\,{\sc i} 617.33~nm in the making. The filter analyzed in detail in this paper is a smaller prefilter for CRISP (32~mm diameter, 26.5 mm clear aperture) which only gives a small FOV reduction compared to the 50-mm filters, since the FOV is currently limited by the polarimetric modulator reused from CRISP. After the etalons, a polarizing beam splitter separates the light in two beams with orthogonal polarization states that are recorded by two cameras (Ximea CB262RG-GP-X8G3). The resulting image scale in the camera focal plane is $0\farcs05$/px.

 \begin{table}
\caption{CRISP2 etalon reflectivities and transmission profile FWHM. The calculation of the FWHM included the broadening induced by the F/140 convergence of the telecentric beam. }              % title of Table
\label{tab:ref}      % is used to refer this table in the text
\centering                                      % used for centering table
\begin{tabular}{c c c c}          % centered columns (4 columns)
\hline\hline                        % inserts double horizontal lines
$\lambda$ [\AA] & $R_{HRE}$ [$\%$] & $R_{LRE}$ [$\%$]& FWHM [m\AA] \\    % table heading
\hline                                   % inserts single horizontal line
5173 & 92.25 & 84.65 & 50\\
5250 & 93.24 & 83.59 & 47\\
5876 & 94.61 & 88.44 & 48\\
5897 & 94.71 & 88.40 & 48\\
6173 & 94.78 & 87.32 & 51\\
6302 & 94.42 & 87.50 & 55\\
6568 & 94.87 & 89.32 & 56\\
7698 & 93.02 & 84.80 & 93\\
7770 & 93.13 & 84.55 & 94\\
8498 & 92.65 & 85.63 & 116\\
8542 & 92.56 & 85.57 & 119\\
\hline                                             %inserts single line
\end{tabular}
\end{table}

\section{Characterization of the CRISP2 FPI system from observational datasets}
\subsection{The CRISP2 transmission profile}\label{sec:tr}
%The theoretical transmission profile of an etalon for a ray with incidence angle $\theta$ relative to the normal is (see, e.g., \citeads{1998A&A...340..569K}):
 {Assuming that there is no absorption by the coatings}, the theoretical  {electric field} transmission profile of an etalon for a ray with incidence angle $\theta$ relative to the normal is \citep[see, e.g.,][]{Hecht2017optics,Born_Wolf_2019}:
\begin{dmath}
  %T(R,C,\theta,\lambda) = \frac{1}{1+\frac{4 R}{(1-R)^2}\sin^2 \left( \frac{2\pi n\cos(\theta)C}{\lambda} \right)},  \label{eq:tr1}
  T(R,C,\theta,\lambda) = \left (\frac{1-R}{1-R\mathrm{e}^{i\psi}}\right )\mathrm{e}^{i\psi/2} = \frac{1}{1-R}\left (\frac{(1-R)\cos(\psi/2) + i(1+R)\sin(\psi/2)}{1+F\sin^2(\psi/2)}\right )\label{eq:tr1}
\end{dmath}
where $\psi=4\pi n\cos(\theta)C/\lambda$, $R$ is the reflectivity ( {intensity reflectance)} of the etalon, $C$ is the cavity separation, $n$ is the refractive index of the cavity ($n=1$ in our case), $F=4R/(1-R)^2$, and $\lambda$ is the wavelength.  {The factor $\mathrm{e}^{i\psi/2}$, commonly ignored in the literature, was recently noted by \citetads{2019ApJS..241....9B}. It is a global phase factor originating from the travel of the first ray that enters the cavity before it is reflected in the second surface. This term has no influence for collimated beams, but it will carry a phase difference between rays with different incidence angles. Arguably, this phase term is very small in comparison with the net phase error accumulated over many reflections inside the cavity, but should be included in the calculations.}

%\sout{The FWHM of the transmission profile given by Eq.~\ref{eq:tr1} can be approximated using the relation (see, e.g., \citeads{2006A&A...447.1111S})}:
%\begin{equation}
%\sout{\mathrm{FWHM} = \lambda^2 / (2 n \mathcal{F} C ),}\label{eq:fwhm}
%\end{equation}
%\sout{where the reflectivity finesse is $\mathcal{F} = \pi \sqrt{R} /(1-R)$. }

As the FPI scans over a spectral line, both etalons are co-tuned to the same wavelength positions.  {Assuming that there are no internal reflections between the two etalons (suppressed here by the LRE tilt and the pupil stop)}, the effective transmission profile at each wavelength is therefore the multiplication of the transmission profiles from each etalon (see, e.g., \citeads{2000A&AS..146..499V}):
\begin{equation}
 T(R_h,C_h,R_l,C_l,\theta,\lambda) = T(R_l,C_l,\theta,\lambda) \cdot T(R_h,C_h,\theta,\lambda),\label{eq:ray}
\end{equation}
where the $h$ and $l$ subscripts correspond to high-resolution and low-resolution etalon parameters. 

A  {rough} simplification is to assume perpendicular incidence with the surface of the etalon, for which we only need to consider the transmission profile along one incidence angle. This simplification can reasonably approximate the real transmission profile when the range of incidence angles of the telecentric beam is very small (usually fulfilled when the F-ratio  {multiplied by the refractive index of the cavity - if different from that of air -} are large). Since non-perpendicular rays decrease the $\cos(\theta)$ term, the profile shifts to shorter wavelengths. In reality, the beam is slowly converging (F/140 for CRISP2, F/165 for CRISP), resulting in a range of incidence angles in the range $[0,1/(2F)]$  {(pupil apodization)}. 

 A better approximation of the real transmission profile can be obtained by  {including the pupil apodization effects by} performing a weighted integral over all incidence angles of the telecentric beam (in this case defining an integration quadrature of  $N_{\mathrm{ray}}$ incidence angles), while assuming axial symmetry around the optical axis.  {Dropping the wavelength dependence of the profile to simplify the notation, the angular integral can be expressed as}:
\begin{dmath}
 T(R_h,C_h,R_l,C_l) = \sum_{i=0}^{N_{\mathrm{ray}}-1} w_i \mathrm{e}^{-ik_4\phi_4}\left \{T(R_l,C_l,\theta_i) \cdot T(R_h,C_h,\theta_i)\right \}, \label{eq:conv}
\end{dmath}
where $w_i$ are normalized integration weights  ($\sum_i w_i = 1$).  {The $\mathrm{e}^{-ik_4\phi_4}$ factor is a focus term that we will discuss in \S\ref{sec:refoc}.}

Since $\theta$ is a radial coordinate, evenly spaced $\theta_i$ are associated with an annular area proportional to $\theta_i$, which would have to be accounted for in the weights, $w_i \propto 2\pi \theta_i\Delta \theta$. However, we can place the rays at angles $\theta_i$, so that $\theta_i^2$ are evenly distributed in $[0,\theta_{\mathrm{max}}^2=1/(2F)^2]$ and then take the square-root of the result:
\begin{equation}
  \theta_i =\sqrt{\frac{1}{(2F)^2} \frac{i}{N_{\mathrm{ray}}-1}}
\end{equation}
where $i$ is the ray number taking values in the range $[0,N_{\mathrm{ray}}-1]$. In this case, the larger contribution from external rings is contained in the ray distribution, which can now have all the same weight. We note that for inclined rays, the transmission profile is shifted towards shorter wavelengths, and therefore the integrated profile from Eq.~\ref{eq:conv} is also slightly shifted compared to perpendicular incidence for an identical cavity separation. The effective shift of the profile can be easily estimated:
\begin{equation}
\Delta \lambda_{\mathrm{peak}} = \left ( \sum_{i=0}^{N_{\mathrm{ray}}-1} w_i \lambda_0 \cos\theta_i \right) - \lambda_0\label{eq:blueshift}
\end{equation}
where $\lambda_0$ is the wavelength where the peak of the perpendicularly incident ray is centered at ($\lambda_0$ is such that $C = m\lambda_0 / 2$, where $m$ is an integer number). 

 {Another effect of the angular integral is that the resulting profile becomes slightly asymmetric.}

%\begin{figure}
%	\centering
%	\includegraphics[width=\columnwidth]{figs/fig_TRs.pdf}
%	\caption{Central lobe of the CRISP2 transmission profile calculated with three different recipes: perpendicular incidence (\emph{ray}, black), telecentric beam without etalon tilt (\emph{conv}, blue) and the full calculation including the tilt of the LRE (\emph{full}, red). \emph{Left}: Transmission profiles. \emph{Right:} peak-normalized transmission profiles, where the blueshift induced by the angular integral has been compensated in the \emph{conv} and \emph{full} calculations for a better comparison of their FWHM and degree of asymmetry.}\label{fig:TRs}
%\end{figure}

The approximation that we just discussed, which we will label as \emph{conv} (for \emph{converging}) hereafter, only requires $N_{\mathrm{ray}} \gtrsim 7$ in order to achieve a sufficiently accurate  angular integral. The downside of this model is that it neglects potential tilts of the etalons to minimize internal reflections. For example, in the case of CRISP and CRISP2, the LRE is tilted by $0.5/F$, inducing an asymmetry in the integrated LRE transmission profile and a slightly lower transmission.

\begin{figure}
	\centering
	\includegraphics[width=\columnwidth]{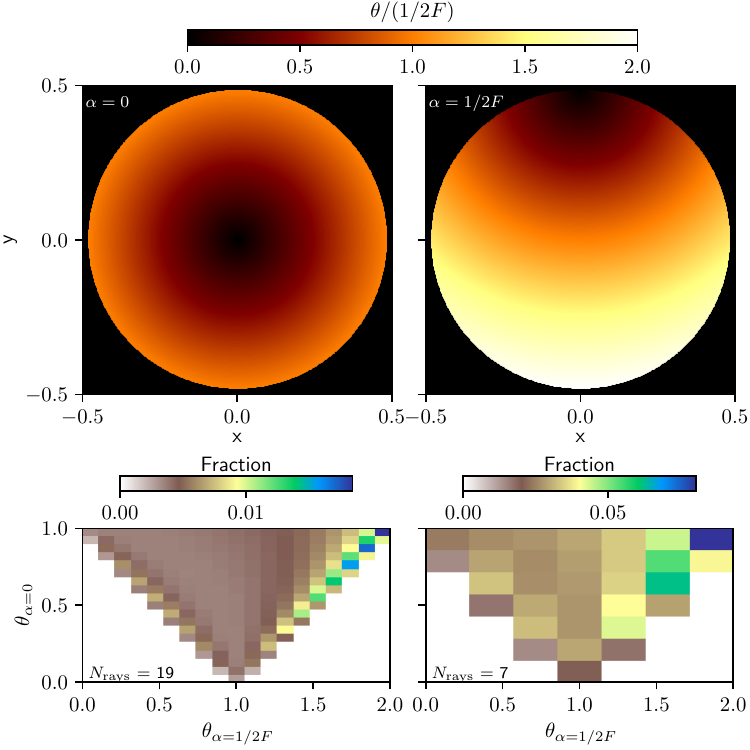}
	\caption{Distribution of inclination angles across the (circular) pupil for a non-tilted case ($\alpha=0$, top-left) and a tilt of $\alpha=1/2F$ in the y-axis (top-right).  {The bottom panels show 2D histograms correlating the angles of the two cases (left: $N_{\mathrm{rays}} = 19$, right:  $N_{\mathrm{rays}} = 7$). This histogram provides the integration weights used to perform the angular integral of the transmission profile over the pupil angles.}}\label{fig:pupil}
\end{figure}

To include the effect of tilting the LRE etalon, the tilt angle in one axis must be added to the angles of the rays in the converging beam.  {This etalon tilt breaks the angular symmetry, making the problem more challenging.}
Figure~\ref{fig:pupil} illustrates the angular distribution across the pupil for a non-tilted and tilted etalons. Clearly, there is  {still} some symmetry in the tilted case that could be exploited to define an analytical integration quadrature, but we find it easier to simply calculate a  {2D} histogram over the angular distributions of the two etalons, and use the central values of the bins as the inclination angles of the different rays.  {We illustrate such histograms in the bottom panels of Fig.~\ref{fig:pupil}, in this case computed using $19$ (left) and $7$ (right) bins. In our tests, using $N_{\mathrm{ray}} \ge 7$ (for each etalon) yields an accurate enough integration, with a maximum error lower than $0.5\%$ relative to a calculation using $101$ rays.} The weight for each bin is given by the value of the histogram at that location. In the following, we will refer to this case as the \emph{full} calculation.  {The tilt in the LRE makes its profile more asymmetric compared to the untilted case, and this asymmetry is propagated to the total transmission profile, mostly notable in the wings of the profile.}

Once the inclination angles and their weights are defined, the effective transmission profile can be obtained with an expression similar to Eq.~\ref{eq:conv}, although the angles are defined for each etalon separately and all permutations must be evaluated in the integral with the corresponding weight.  {The intensity transmission profile is calculated as:}
\begin{equation}
\hat{T} = T \cdot T^\dagger,\label{eq:Itr}
\end{equation}
 {where $T^\dagger$ is the complex conjugate of $T$. When phase errors are small, the resulting transmission profile that originates from using Eq.~\ref{eq:tr1}, \ref{eq:conv} and \ref{eq:Itr} are similar to those resulting from integrating individual intensity transmission profiles, although mathematically this is not strictly the case since:}
\begin{equation}
\sum_i(T\cdot T^\dagger) \ne \bigg(\sum_i T \bigg) \cdot \bigg(\sum_i T \bigg )^\dagger. \label{eq:int}
\end{equation}

%CRISP has a higher F-ratio than CRISP2 (F/165 vs F/140). Therefore, for CRISP the broadening effect from the angular integral in Eq.~\ref{eq:conv} is smaller and the transmission profile FWHM predicted  {by the profile of a single ray} at 617.3~nm is closer to reality (only $\sim12\%$ lower) than for CRISP2, for which the peak transmission  is $\sim26\%$ narrower than the real one. We emphatically discourage the utilization of Eq.~\ref{eq:tr1} (assuming perpendicular incidence) with CRISP and CRISP2 as it yields an unrealistically narrow transmission profile. This effect is also illustrated in Fig.~\ref{fig:TRs}, where we show the central part of the profile calculated with each of the three methods. While the \emph{conv} and \emph{full} calculations are almost identical, the \emph{ray} calculation is much narrower.

In this study, we have chosen to express the cavity separation errors of the LRE ($\Delta C_{\mathrm{LRE}}$) relative to the HRE ones as:
 \begin{flalign*}
  C_{\mathrm{HRE}}(x,y) &= C_{\mathrm{HRE}}^0 + \Delta C_{\mathrm{HRE}} (x,y)\\
 C_{\mathrm{LRE}} (x,y)&= C_{\mathrm{LRE}}^0 + \Delta C_{\mathrm{LRE}}(x,y) + \Delta C_{\mathrm{HRE}}(x,y)\cdot r_c,
 \end{flalign*}
 where $C_{\mathrm{HRE}}$ and $C_{\mathrm{LRE}}$ are the FOV-dependent cavity separations. In this way, the LRE is always co-tuned with the HRE unless $\Delta  C_{\mathrm{LRE}} \ne 0$. As a result, the cavity error map of the LRE does not need to be compensated for the HRE cavity errors after the fitting. %The reason for this choice will become more obvious in \S\ref{sec:lremodel}, where we describe the model to fit the LRE parameters.

 \subsection{System optimization by re-focusing} \label{sec:refoc}
 {\citetads{2006A&A...447.1111S} performed an extensive analysis of the phase error amplification function. The phase error amplification function has a quadratic dependence across the pupil, mimicking a de-focus term. However, it also varies in strength across the transmission profile. Therefore, optimally refocusing the instrument will reduce, but cannot eliminate, these phase errors in the system. Although the analysis was mostly described in a context of improving image quality, refocusing also improves the overall transmission of the instrument.}

 {The refocusing term can be trivially modeled using a de-focus Zernike term $\mathrm{e}^{-ik_4\phi_4}$, where $k_4$ is a constant that captures the amount of de-focus, $\phi_4 = \sqrt{3}(2r-1)$, and $r$ is the normalized radius in the pupil for a given ray. The constant $k_4$ can be optimized once to achieve maximum transmission. In practice, this term is automatically accounted for when the instrument is focused on the optical table. }

\begin{figure}
	\centering
	\includegraphics[width=\columnwidth]{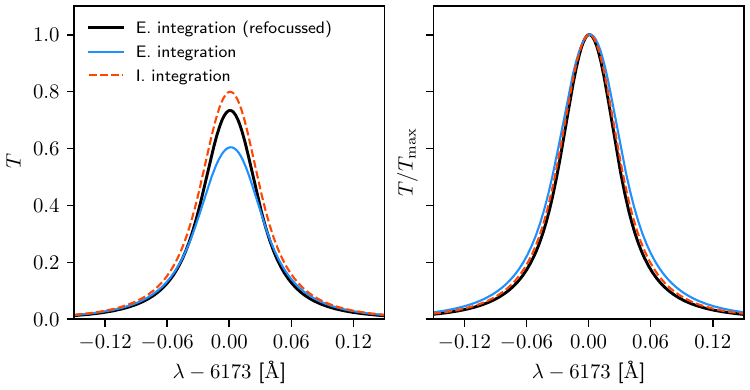}		
	\caption{ {CRISP-2 transmission profiles calculated including instrument refocus (black), no refocus (blue), and by performing the angular integral directly on the intensity transmission profiles (dashed-red). The profiles in the right panel are peak normalized.}}\label{fig:refoc}
\end{figure}

 {Figure~\ref{fig:refoc} shows a comparison of profiles calculated with and without the re-focus term, and one profile calculated by direct integration of the intensity transmission profile (ignoring phase errors). Without re-focusing (blue curve), the peak transmission is very low ($76 \%$ of the phase-error-free case in dashed-red) and the profile is wider, whereas with the re-focusing term it is possible to reach $92\%$ of that value and the resulting profile is only slightly narrower. We suspect the latter is due the approximation introduced in Eq.~\ref{eq:int} for the integration of the intensity transmission profile.}

 {Modern computers are very efficient in operations with complex numbers and our implementation of the electric field transmission profile angular integral is only $\sim50-60\%$ slower than performing the intensity transmission profile one using real numbers.}
 
\subsection{The prefilter transmission curve}
The prefilter curve can be modeled by an analytical formula:
\begin{equation}
P(\lambda) = \frac{p_{\mathrm{g}}}{1 + \left ( 2 \frac{\lambda - p_{w_{0}} }{ p_{\mathrm{fwhm}} } \right)^{2 p_{\mathrm{ncav}}} } \left \{1 +  w_{\mathrm{apod}}(p_0 \Delta\lambda+ p_1 \Delta\lambda^2 + p_2 \Delta\lambda^3) \right\}, \label{eq:pref}
\end{equation} 
where $p_{\mathrm{g}}$ is a gain factor, $p_{w_{0}}$ is the central wavelength of the prefilter, $p_{\mathrm{fwhm}}$ is the prefilter full-width-half-maximum, $p_{\mathrm{ncav}}$ is the number of cavities of the prefilter, $\Delta\lambda = \lambda - p_{w_0}$, and $p_0$, $p_1$ and $p_2$ are the coefficients of a polynomial to account for deviations in the analytical prefilter shape ( {including asymmetries and spatial movement of fringes with wavelength}). Similar expressions were used by \citetads{2011A&A...534A..45S} and \citetads{2015A&A...573A..40D} to describe the shape of interference prefilters, but in our case the polynomial component is defined relative to the prefilter central wavelength in order to have a shift-invariant prefilter shape. We have also introduced an apodization window (in the spectral dimension) that multiplies the polynomial component ($w_{\mathrm{apod}}$) to avoid negative values in the far wings of the prefilter. The latter is defined relative to the central wavelength and FWHM of the prefilter:
\begin{equation}
w_{\mathrm{apod}} = \frac{1}{4}\bigg\{(1 + \tanh(\Delta \lambda_1)) \cdot (1-\tanh(\Delta \lambda_2))\bigg\},
\end{equation}
where,
\begin{flalign}
\Delta \lambda_1 &= \frac{\pi}{2}\left( \frac{\lambda - p_{w_0}}{a_{\mathrm{scl}}p_{\mathrm{fwhm}}}  +  a_{\mathrm{off}}\right),\\
\Delta \lambda_2 &= \frac{\pi}{2}\left( \frac{\lambda - p_{w_0}}{a_{\mathrm{scl}}p_{\mathrm{fwhm}}}  -  a_{\mathrm{off}}\right).
\end{flalign}
We have chosen $a_{\mathrm{off}}=5$ and $a_{\mathrm{scl}}=0.5$ for our Alluxa prefilters (for which $p_{\mathrm{fwhm}}\approx 5$~\AA\ and $p_{\mathrm{ncav}}\approx 2$), but other combinations could work better with other prefilters. These parameters regulate the width and decay of the apodization window and they are kept constant during the fit. We also note that modern prefilters can have many dielectric coating layers, and the $p_{\mathrm{ncav}}$ parameter no longer translates to the physical number of cavities but rather it is used here as a free parameter to fit the prefilter shape. 

\begin{figure}
	\centering
	\includegraphics[width=\columnwidth]{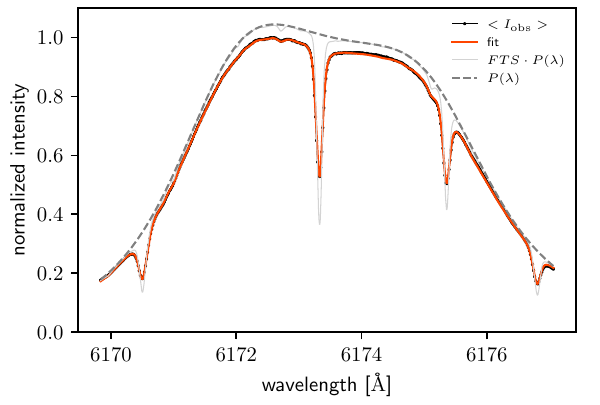}		
	\caption{Observed mean intensity (black) and the derived fit (red) in the 617.3~nm spectral window. The inferred prefilter curve is plotted with a dashed gray line. The light-gray spectrum represents the FTS atlas multiplied by the prefilter curve. The fitted curve is the result of the FTS spectrum multiplied with the prefilter curve and convolved with the nominal CRISP2 transmission profile.}\label{fig:hre_data}
\end{figure}

\subsection{Datasets and model for the HRE and prefilter characterization}\label{sec:hremodel}
The model used to fit the observed spectrum at any $(x,y)$ location in the FOV is given by the following expression:
\begin{equation}
I_{\mathrm{obs}} (\lambda) = \frac{\int_{\lambda_0}^{\lambda_1}\left\{ I_{\mathrm{FTS}} (\lambda') \cdot P(\lambda') \right\} \cdot \hat{T}_{\mathrm{CRISP2}}(\lambda'-\lambda)d\lambda'}{\int_{\lambda_0}^{\lambda_1}d\lambda'},\label{eq:hremodelc}
\end{equation}
 {where $\hat{T}_{\mathrm{CRISP2}}$ is, a priori, wavelength dependent. However, within the spectral range covered by the prefilter, such variation is very small and the transmission profile could be assumed to be constant. We have performed test calculations (see Appendix~\ref{ap:conv}) showing that the maximum error made in the derived prediction is $\sim1\%$ of the peak intensity, and generally much lower than that. Therefore, we can express the Eq.~\ref{eq:hremodelc} as a convolution with a wavelength invariant transmission profile, in our case evaluated in the center of the prefilter:}
\begin{equation}
I_{\mathrm{obs}} (\lambda) = \left\{ I_{\mathrm{FTS}} (\lambda) \cdot P(\lambda) \right\} \circledast \hat{T}_{\mathrm{CRISP2}}(\lambda),\label{eq:hremodel}
\end{equation}
 {where $\circledast$ represents a convolution. This form is computationally more efficient as the transmission profile does not need to be recalculated for each wavelength point of the model (typically a few hundreds).}
The model has nine free parameters: seven prefilter parameters from Eq.~\ref{eq:pref} and the reflectivity and cavity separation of the HRE from Eq.~\ref{eq:ray} {(as used in Eq. \ref{eq:conv})}. 

When the two etalons are co-tuned, the transmission profile is set by  {the angular integral of the product of the profiles ($\hat{T}_{\mathrm{CRISP2}}$) from both etalons (Eq.~\ref{eq:conv}). Although both etalons contribute to the exact shape of the resulting transmission profile, the latter is more strongly modulated by the HRE parameters than from the much broader LRE}. Therefore, it is reasonable to propose that by acquiring a co-tuned scan of solar spectra, one could  {constrain} the reflectivity and the cavity separation of the HRE if the input solar spectrum is known a-priori for each pixel (in this case the quiet-Sun mean spectrum at disk center). We exploit the fact that the spatio-temporal average of the quiet-Sun solar spectrum at disk center can be assumed to be statistically the same at any given time (e.g., \citeads{1981A&A....96..345D}). Therefore, datasets acquired over a sufficient amount of time while the telescope pointing is moving around the disk center position should yield the same input spectrum in each of the pixels. Variations across the FOV must therefore originate from variations in the FPI etalons, fringes, and the prefilter. We use the Fourier Transform Spectrometer solar atlas acquired at McMath Solar Telescope of \citet{1984SoPh...90..205N}, hereafter the FTS atlas, as  {the} template quiet-sun average spectrum ($I_{\mathrm{FTS}}$) to compare with the CRISP2 data described below.

The data considered for this part consists of a long-range spectral scan acquired in flat-field mode, so that the imprint of the prefilter in the observed spectra is properly captured. The resulting dataset is a 3D cube with two spatial and one spectral dimension, $(n_x,n_y,n_\lambda)$, where each spatial pixel has a realization of the observed spectrum. The wavelength step size was $\delta \lambda = 16$~m\AA.

%If the etalons were perfectly flat, with strictly constant reflectivity, and the prefilter transmission profile was constant across the FOV, all pixels should contain an identical spectrum. Obviously that is not the case as the surface of the etalons and the prefilter cannot be made perfectly flat, and measurable variations are present across the FOV for most model parameters. 

Figure~\ref{fig:hre_data} depicts the observed mean intensity (averaged over the central part of the FOV) and the best fit using the FTS atlas. The FTS has been acquired at much higher spectral resolution than that available in CRISP2 observations. We note that the observed spectrum in that figure (the spatial average over the FOV) is broader than in each of the pixels due to the shifts induced by cavity errors across the FOV. 

The method presented in this paper is expected to work and yield sensitive results for the etalon reflectivity determination when the atlas spectrum, convolved with the FPI transmission profile, shows a significant difference compared to the unconvolved one. For example, in a hypothetical case in which the transmission profile is much narrower than the spectral lines, the effect will likely be very weak and the sensitivity to variations in the FPI reflectivities will be low. We have illustrated this effect in Fig.~\ref{fig:hre_data} by also including the product of the FTS atlas with the prefilter curve (but not convolved with the FPI transmission profile). A similar model was utilized by \citet{delacruz2010} and \citet{2013A&A...553A..63S}. The former report was never published and the method was barely mentioned in the preparation of the flat-fields in \citet{2013A&A...553A..63S}. Both cases utilized the \emph{conv} transmission profile recipe,  {performing a computationally cheaper angular integration of the intensity transmission profile}.

Model simplifications can be adopted in order to speed up the process, by trading some physical accuracy. For example, by only including the upper part of the prefilter scan (e.g., $[6173.0-6173.5]$~\AA), the number of spectral points is largely reduced and we can use a much simpler description for the prefilter, likely only containing $p_{\mathrm{g}}$ and a linear term from the polynomial component ($P_0$). In this case, we would only characterize the HRE etalon parameters, leaving aside the full description of the prefilter. However, the latter would only be appropriate when the prefilter is properly damping the secondary transmission peaks at $\pm 1$ times the free spectral range (FSR). 

\subsection{Datasets and model for the LRE characterization}\label{sec:lremodel}
\begin{figure}
	\centering
  \includegraphics[width=\columnwidth]{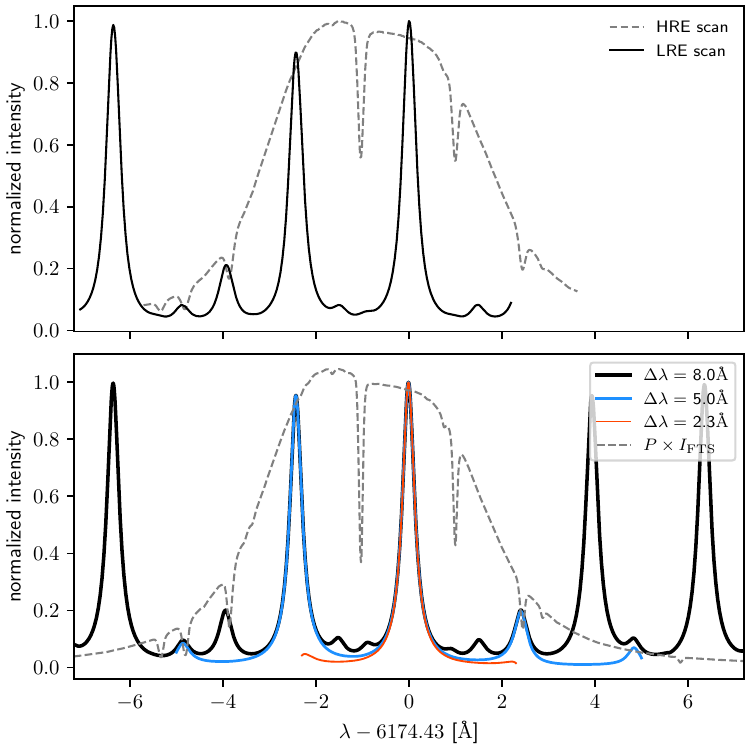}
	\caption{
	\emph{Top:} Observed spatially averaged LRE (black) and HRE (dashed grey) spectral scans. \emph{Bottom:} 
	Simulated LRE scans, only including the central lobe of the HRE and LRE (red), including the secondary HRE lobes at $\pm 1\times\mathrm{FSR}$ (blue), and including the secondary HRE peaks at $\pm 2\times\mathrm{FSR}$ and the LRE peaks at $\pm 1\times\mathrm{FSR}$ (black). The HRE profile has been multiplied by an analytical estimate of the prefilter and the FTS atlas (shown in dashed grey).}\label{fig:lre_data}
\end{figure}

The profile of the LRE can be \emph{sampled} by parking the HRE in a continuum wavelength close to the central wavelength of the prefilter, and scanning in the spectral direction with the LRE:
\begin{equation}
%I_{\mathrm{obs}} (\lambda)^{\mathrm{LRE}} = \left\{ I_{\mathrm{FTS}}(\lambda) \cdot P(\lambda) \cdot \hat{T}_{\mathrm{HRE}}(\lambda) \right\} * \hat{T}_{\mathrm{LRE}}(\lambda),\label{eq:lremodelc}
I_{\mathrm{obs}} (\lambda)^{\mathrm{LRE}} = \frac{\int_{\lambda_0}^{\lambda_1}\left\{ I_{\mathrm{FTS}}(\lambda') \cdot P(\lambda') \right\} \cdot \hat{T}_{\mathrm{CRISP2}}(\lambda'-\lambda,\lambda')d\lambda'}{\int_{\lambda_0}^{\lambda_1}d\lambda'},\label{eq:lremodelc}
\end{equation}
 {where $\hat{T}_{\mathrm{CRISP2}}(\lambda'-\lambda,\lambda')$ is calculated with the HRE placed at a constant wavelength, while the LRE scans in the spectral direction (hence the double $\lambda$ dependence in the notation). The position of the LRE can be regulated by adding an offset to the cavity error value. We cannot directly assume that the transmission profile is wavelength invariant. If we neglect the slow wavelength variation of the transmission profile and we assume that the angular integral can be performed separately for each etalon, such that:}
\begin{eqnarray*}
	\hat{T}_{\mathrm{HRE}} &=& \sum_{i=0}^{N_{\mathrm{ray}}-1} w_i \mathrm{e}^{-ik_4\phi_4} T(R_H,C_H,\theta_i), \\
	\hat{T}_{\mathrm{LRE}} &=& \sum_{i=0}^{N_{\mathrm{ray}}-1} w_i \mathrm{e}^{-ik_4\phi_4} T(R_L,C_L,\theta_i),
\end{eqnarray*}
 {we can re-write the model as a convolution:}
\begin{equation}
I_{\mathrm{obs}} (\lambda)^{\mathrm{LRE}} = \left [ \left ( I_{\mathrm{FTS}}\cdot P\cdot \hat{T}_{\mathrm{HRE}}\right\} \circledast \hat{T}_{\mathrm{LRE}}\right ] (\lambda).\label{eq:lremodel}
\end{equation}
 {We have tested the validity of these two assumptions in Fig.~\ref{fig:convwave} (see right panel). Our calculations show that the maximum difference between the predictions from Eq.~\ref{eq:lremodelc} and \ref{eq:lremodel} is very small ($\pm 0.2\%$ relative to the peak intensity) in the central part of the profile. Therefore we consider this approximation adequate for the purposes of this study.}

Given that the HRE profile is much narrower than the LRE profile, their convolution yields a train of LRE profiles centered at the locations of the HRE transmission peaks. The strength of the secondary peaks is modulated by the transmission  of the prefilter. If the latter is not strictly symmetric and the HRE is not centered at the central wavelength of the prefilter, we should expect asymmetries in the peak amplitude of secondary transmission lobes (see, e.g., Fig.~\ref{fig:lre_data}). %Figure~\ref{fig:lre_data} (upper panel) represent the spatial average of such a scan performed with the CRISP instrument, where we have also plotted the mean prefilter curve. 

The weak transmission peaks at approximately $\pm 1.7$~\AA\ originate from the first FSR peak of the LRE aligning with the second FSR peak of the HRE on the opposite side of the prefilter curve (see the lower panel in Fig.~\ref{fig:lre_data}). This means that in order to model those peaks close to the center, our model must include at least the first FSR peaks of the LRE (at $\Delta \lambda \approx \pm 6.5$~\AA) and reach two FSR peaks for the HRE. Since the determination of the prefilter in the far wings is less accurate than in the central part of the prefilter, and fitting one lobe should suffice to determine the cavity and reflectivity errors, we have opted for only fitting the central peak, although the dataset  {and model} cover a much larger spectral range  {to properly reproduce the wings of the central lobe}.  

The prefilter parameters and HRE parameters that also appear in Eq.~\ref{eq:lremodel} are derived in the HRE fits for each pixel in the FOV, and kept constant for the LRE fits. But we allow the prefilter gain to change as the LRE and HRE scans are not strictly simultaneous and the intensity level might have changed. The model has therefore three free parameters per pixel (the gain factor, LRE cavity separation and LRE reflectivity).

\begin{figure}
	\centering
	\includegraphics[width=\columnwidth, trim=0 0.3cm 0 0 clip]{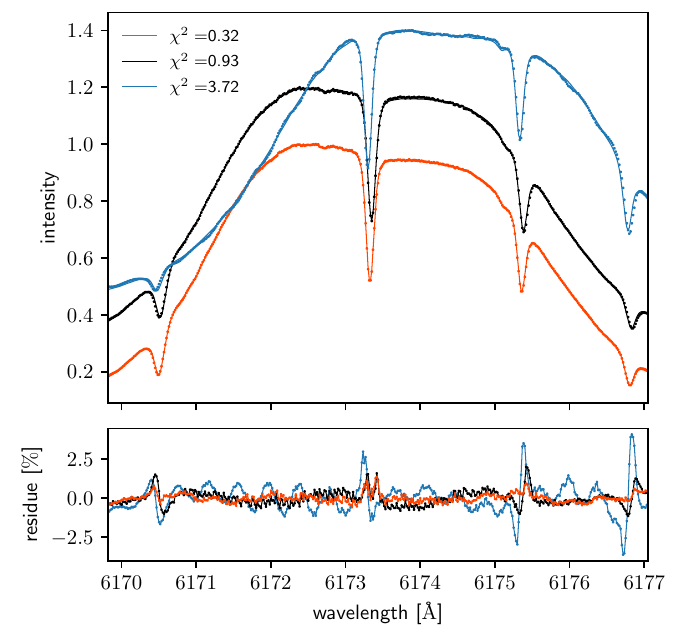}
	\caption{Observed spectrum (dots) and best-fit (solid line) from three locations in the FOV. An offset of $\pm0.2$ was applied to the blue and red curves to improve readability. The locations of these points are indicated in panel (a) of Fig.~\ref{fig:hreres} using the same color coding. The residues are normalized by the peak observed intensity of each curve.}\label{fig:hrefits} 
\end{figure}

\begin{figure*}
	\centering
	\includegraphics[width=\textwidth]{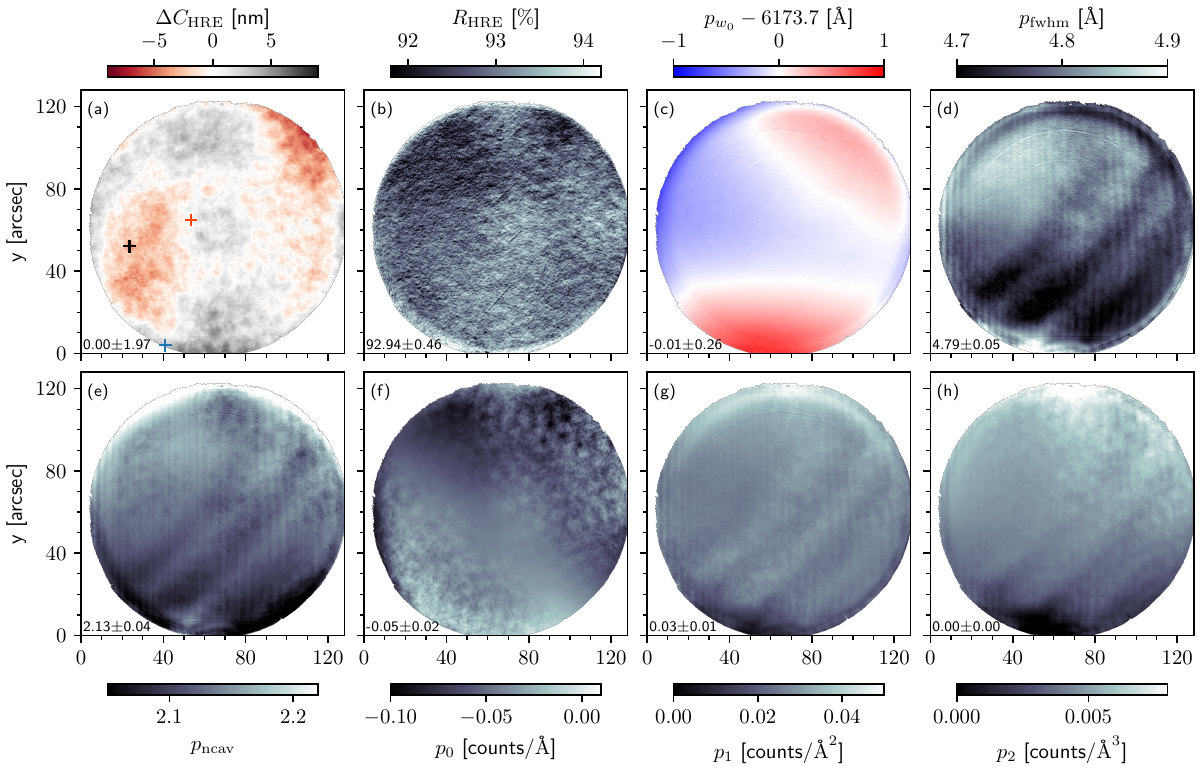}
	\caption{Inferred parameter maps from the prefilter and HRE model. (a) is the HRE cavity error map, (b) is the HRE reflectivity map, (c) is the prefilter central wavelength, (d) is the prefilter FWHM, and (e)-(g) are the coefficients of the polynomial prefilter components. The crosses indicated in (a) correspond to the fits shown if Fig.~\ref{fig:hrefits} using the same color coding.}\label{fig:hreres}
\end{figure*}

\subsection{Code implementation}
We have implemented a Levenberg--Marquardt algorithm \citep{levenberg1944method,marquardt1963algorithm} to fit the data. 
The code is written in C++ with OpenMP parallelization. We have created a Python interface for all routines\footnote{These codes along with documented examples are publicly available on \url{https://github.com/jaimedelacruz/pyFPI}}. The FPI functions are implemented using object-oriented programming, so that many quantities are only calculated once upon initialization of the class, and kept for subsequent calculations. All three transmission profile approximations (see \S\ref{sec:tr}) and their analytical derivatives  are available through the Python interface. We briefly describe the calculation of the analytical derivatives of the transmission profile and the prefilter curve in Appendices~\ref{ap:der} and \ref{ap:prefvar}. Our code makes use of the Eigen-3 linear algebra library \citep{eigenweb}.

%The Levenberg-Marquardt algorithm was also implemented by us in C++ in a very generic way, so this implementation can be used for other projects / models in a very straight forward way. In a similar fashion to the older MPFIT routines \citepads{2009ASPC..411..251M}, it allows setting parameter range limits and fixing the value (or not) of the different model parameters, allowing for easy experimentation with the complexity of the model. The latter proves to be very useful when the observations sample a smaller spectral range or if (hypothetically) degeneracies between the model parameters appear in a given spectral window.

\section{Results}\label{sec:res}
The calculations have been performed as follows:
\begin{enumerate}
	\item Initial estimates of the HRE and LRE cavity maps are calculated by fitting a parabola to the core of a spectral line (HRE) or the maximum of the intensity peak (in the LRE dataset). The wavelength shifts are converted to cavity separation $(\Delta \lambda / \lambda = - \Delta C/ C)$. The initial LRE cavity map is compensated for the HRE cavity map. \label{item:init}
	\item The HRE and prefilter parameters are fitted, while the LRE initial cavity map is kept constant.\label{it:hre}
	\item The LRE parameters are fitted once the HRE and prefilter parameters are known. The prefilter curve and HRE parameters are used to generate the LRE model.
	\item The HRE parameters from step \ref{it:hre} are refined with the new estimate of the LRE cavity map and reflectivities.\label{it:re}
\end{enumerate}
The LRE cavity map initialization in step~\ref{item:init} allows obtaining a value of the HRE reflectivity that is not affected by the assumption of having the LRE tuned at the reference wavelength or always co-tuned with the HRE. The HRE parameters are re-calculated in step~\ref{it:re} with the final LRE cavity and reflectivity maps, initializing the fitting from the results of step~\ref{item:init}.

\subsection{HRE parameter maps}\label{sec:hreres}
We have optimized the parameters of the model described in Eq.~\ref{eq:hremodel} for each pixel of the observed FOV, obtaining a 2D map for each of the model parameters. We first performed a fit on the mean spectrum over the entire FOV, and used the resulting prefilter parameters as initial values for the 2D fits. An absolute wavelength calibration was obtained in this step, {which is used to perform the translation from digital etalon offset units to a calibrated wavelength axis. This way we can also know precisely where the LRE is parked in the LRE scan.}

Figure~\ref{fig:hrefits} illustrates the spectra from three locations in the FOV and the corresponding best-fit for each of them. We chose these points randomly, but in locations with remarkably different values of $\chi^2$, so that the reader can assess the quality of the fits in regions with low (red), medium (black) and large (blue) values of $\chi^2$. In our opinion, even the example with the highest $\chi^2$ value poses an acceptable fit as the spectral lines width and shape and the overall prefilter curve are reproduced. Most of the discrepancies indicate small-scale structure (in the spectral direction) in the prefilter curve that cannot be fully captured by our model parametrization. 

The resulting model parameters are shown in Fig.~\ref{fig:hreres}. 
%We have not shown 
We do not show maps of $p_{\mathrm{g}}$ as it has no scientific value in this study. The HRE cavity error map (a) has a standard deviation of $2$~nm (not to be confused with the rms of the wavelength shifts that they induce on the spectrum). The HRE reflectivity map has a standard deviation of $0.46\%$. Both quantities were measured over the entire circular FOV and are in relatively good agreement with the specs of the instrument, although the mean $R_{\mathrm{HRE}}$ is $1.8\%$ lower than the nominal expected value from factory measurements. This discrepancy could be explained by a small (residual) tilt angle in the HRE that is compensated by our model as a reduced reflectivity.

The $617.3$~nm prefilter\footnote{This prefilter was manufactured by Alluxa.} is slightly tilted in order to avoid reflections / interference patterns. This tilt shifts the center of the passband to shorter wavelengths  {and can affect the inferred FOV-dependent central-wavelength} (e.g., \citeads{2011A&A...533A..82L}). Additionally, there is a FOV variation of approximately $\pm1$~\AA\ ($0.26$~\AA\ RMS). The prefilter FWHM also varies across the FOV covering a range of approximately $\pm0.3$~\AA\ centered around $4.8$~\AA\ ($0.04$~\AA\ RMS). The coefficients of the polynomial component capture the level of asymmetry (non-zero values in $p_0$) and deviations between the predicted shape given by $1/(1+q^{2p_{\mathrm{ncav}}})$ and the observations. 

The reconstructed prefilter maps show spatial correlations between some of the parameters. They seem to mostly capture fringes, but we note that their imprint in those parameters does not have a large amplitude. We have also tried fixing some of the parameters to a constant value, but the quality of the fits became worse. We decided to let the model parameters absorb the influence of fringes (which are in the data) and obtain the best fit we could, so that the cavity error and reflectivity estimates would not be affected by errors in the prefilter estimation.

%We also noticed that the use of $\Delta \lambda  = \lambda - p_{w_{0}}$ instead of a fixed reference wavelength ($\lambda_{\mathrm{ref}}$), really helped improving the quality of the fits. The latter is probably due to the fact that the effect of the polynomial component becomes shift-invariant compared to using a fixed reference wavelength.

\begin{figure}[!ht]
	\centering
	\includegraphics[width=\columnwidth]{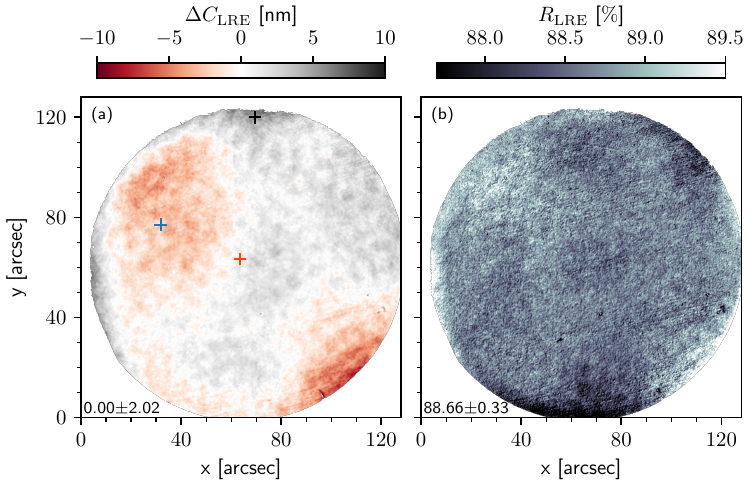}
	\includegraphics[width=\columnwidth]{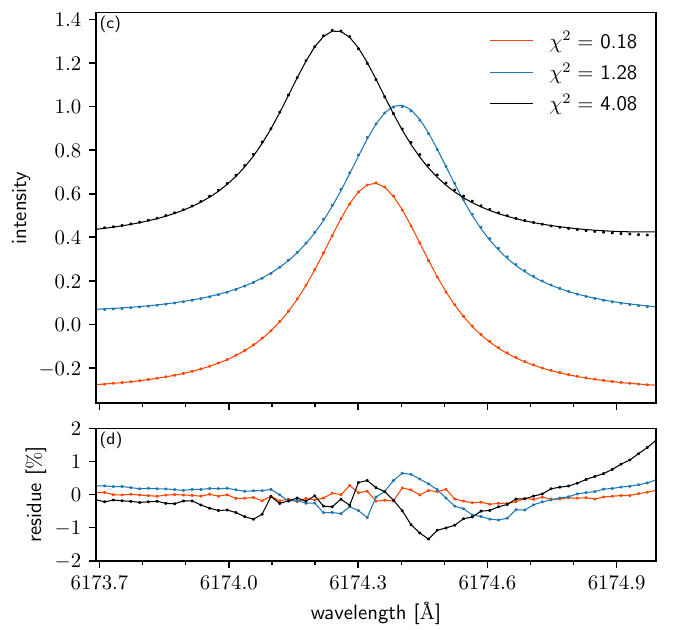}
	\caption{Inferred parameter maps from the LRE model fit and three fit examples. (a) the LRE cavity error map, (b) the LRE reflectivity map, and (c) illustrates three observed LRE scan spectra (dots) and the corresponding fits (solid line). The locations of these spectra in the FOV are marked in panel (a) with cross markers using the same color coding than in panel (c). A vertical offset of $\pm0.35$ was applied to the red and black curves to improve readability.  {The residues are normalized by the peak observed intensity of each curve.}}\label{fig:lreres}
\end{figure}

\subsection{LRE parameter maps}
The LRE model has only three parameters, but (as illustrated in Fig.~\ref{fig:lre_data}) it generally requires knowledge of the spatially resolved prefilter curve and the HRE parameters that were derived in \S\ref{sec:hreres}. Those are passed as a fixed input to the LRE fitting routines, but the gain factor of each pixel is fitted along with the LRE cavity error map and LRE reflectivity map. We note that in order to properly fit the data, accurate and consistent wavelength calibration of all datasets is required, so that the prefilter curve derived with the HRE dataset can be placed at the correct line positions of the LRE scan. {The wavelength calibration performed with the FTS atlas using the mean spectrum allows to accurately perform the translation from digital units to absolute wavelength, under the assumption that the etalons wavelength offset calibration remains constant during the entire data acquisition time. In our experience, this calibration is very stable within many hours or even days.}

The derived 2D maps and fit examples are shown in Fig.~\ref{fig:lreres}. We show fit examples that were chosen to show a low $\chi^2$ (red), a pixel with a $\chi^2$ very close to the mean $\chi^2$ over the FOV (blue) and a large value of $\chi^2$ (black).  Overall, the quality of the fits is very good.  In early attempts including the entire observed LRE spectral range, we noticed that most of the fit discrepancies originate from errors in the secondary transmission peaks  {(even the small ones very close to the central one)}, in the form of an amplitude error. The latter are modulated by the prefilter curve, which likely points to inaccuracies in the prefilter estimation in the very far wings. We note that the prefilter was estimated with the HRE dataset, which in our case is less extended in wavelength coverage than the LRE scan, and therefore, the prefilter in the outer points is not necessarily well constrained by the data, but rather assumed to follow the analytical description in Eq.~\ref{eq:pref}. 

In any case, we have opted for only fitting the central lobe in this part, which is very well reproduced by the model in all cases, even at locations with larger $\chi^2$ values.  {The model itself includes the full range, but only the central lobe is used in the computation of $\chi^2$. Including the full range is important to properly model the wings of the central lobe because there is a significant contribution to the intensity from transmission peaks aligning in the wings of the prefilter}. At 617.3~nm, the mean LRE reflectivity is $R_{\mathrm{LRE}}=88.66 \%$ and the RMS variation of the cavity errors is $\sigma_{C_{\mathrm{LRE}}}=2$~nm. 

\subsection{Simplified models}\label{sec:simp}
The calculations used to  {derive} the parameter maps shown in Figs.~\ref{fig:hreres} and \ref{fig:lreres} were performed using the full calculation of the transmission profile. Furthermore, we made sure that the secondary peaks located at $\pm1\times \mathrm{FSR}$ were included in the calculation of the HRE model. Given the large size of the CRISP2 images ($2560\times 2560~\mathrm{pixels}^2$) and the large number of spectral points included in the model, the calculation of the system parameters with this approach is costly from a computational perspective.
\begin{figure*}
	\centering
	\includegraphics[width=\columnwidth]{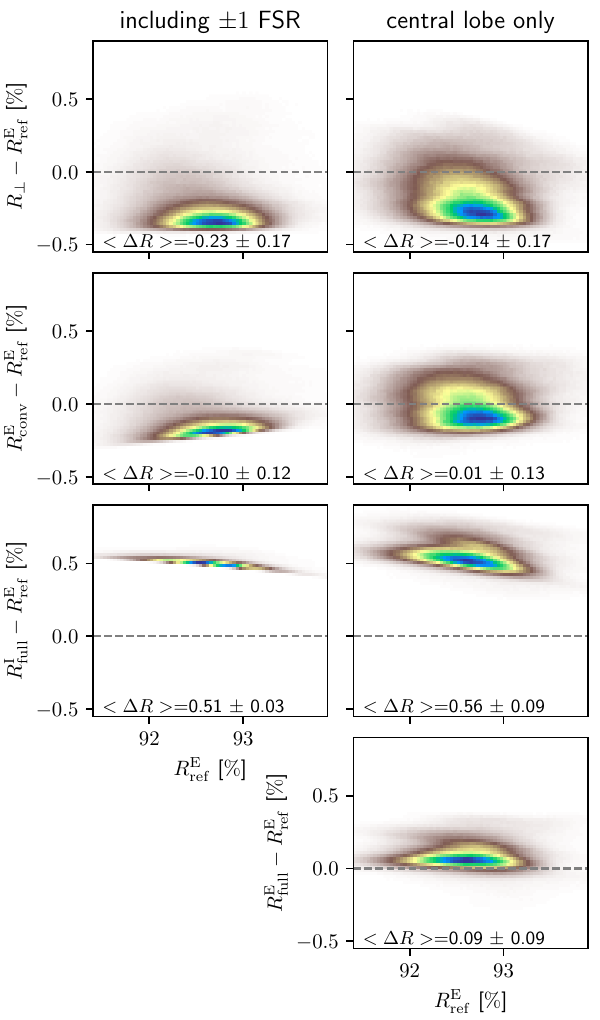}\includegraphics[width=\columnwidth]{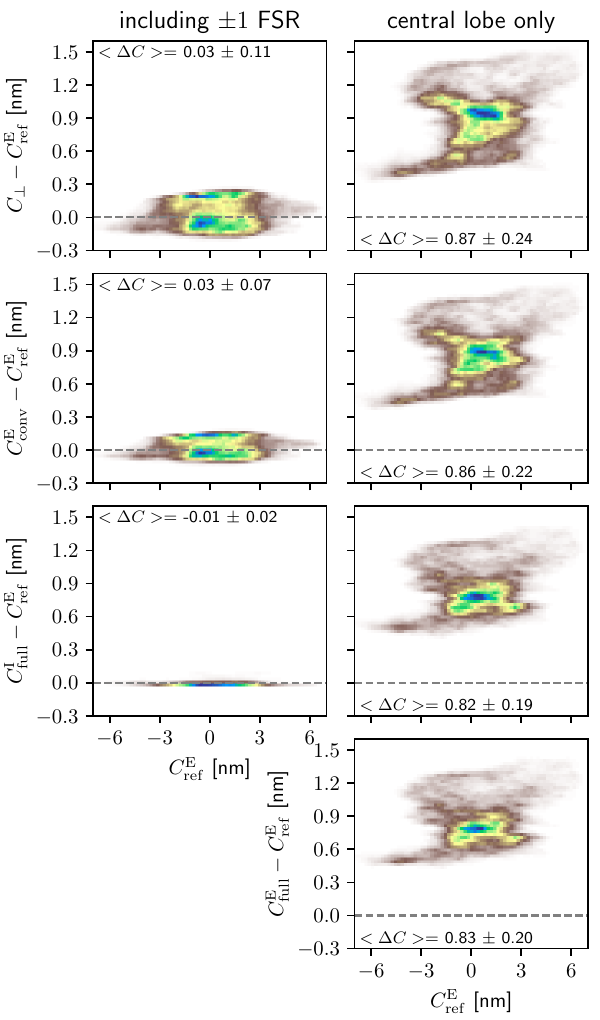}
	\caption{ {2D histograms comparing the inferred HRE reflectivity and cavity errors to the reference case ($R_x^y - R_{\mathrm{ref}}^{\mathrm{E}}$ vs $R_{\mathrm{ref}}^{\mathrm{E}}$ and $C_x^y - C_{\mathrm{ref}}^{\mathrm{E}}$ vs $C_{\mathrm{ref}}^{\mathrm{E}}$). The super-index C indicates that the angular integral was performed over the electric field transmission profile, whereas the I super-index corresponds to a direct integral of the per-ray intensity transmission profile. For each parameter, the results on the left column were calculated using a broad spectral coverage that includes the first pair of secondary transmission lobes. The results on the right column were calculated with a truncated dataset and model, where only the central lobe of the transmission profiles was considered. From top to bottom, the transmission profiles were calculated assuming perpendicular incidence, the \emph{conv} approximation, the \emph{full} integration of the intensity transmission profile, and the \emph{full} (complex) calculation. The reference case is the full (complex) calculation with the broad spectral range.}}\label{fig:hrehist}
\end{figure*}

%\begin{figure}[!ht]
%	\centering
%	\includegraphics[width=\columnwidth]{figs/fig_histo_cmaps.pdf}
%	\caption{2D histograms comparing the inferred HRE cavitymaps to the reference case ($R_x^y - R_{\mathrm{ref}}^{\mathrm{C}}$ vs $R_{\mathrm{ref}}^{\mathrm{C}}$). The super-index C indicates that the angular integral was performed over the electric field transmission profile, whereas the I super-index corresponds to a direct integral of the per-ray intensity transmission profile. The results on the left column were calculated using a broad spectral coverage that includes the first pair of secondary transmission lobes. The results on the right column were calculated with a truncated dataset and model, where only the central lobe of the transmission profiles was considered. From top to bottom, the transmission profiles were calculated assuming perpendicular incidence, the \emph{conv} approximation, the \emph{full} integration of the intensity transmission profile, and the \emph{full} (complex) calculation. The reference case is the full (complex) calculation with the broad spectral range.}\label{fig:lrehist}
%\end{figure}

\begin{figure*}[!ht]
	\centering
	\includegraphics[width=\textwidth]{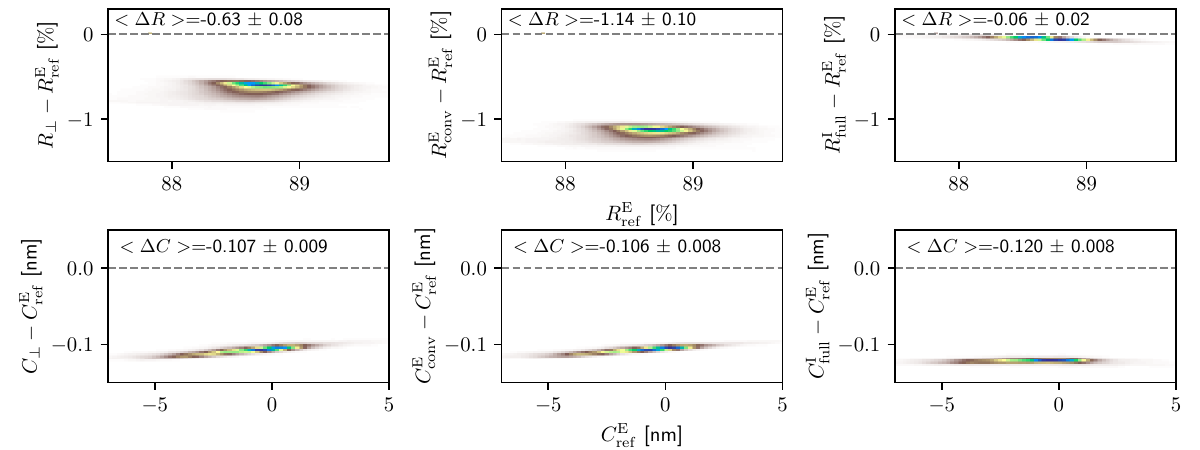}
	\caption{ {2D histograms comparing the inferred LRE reflectivity and cavity error maps to the reference case. The reference case is set by the \emph{full} complex calculation. The upper row shows the results for the reflectivity whereas the bottom row corresponds to the cavity errors. The super-index C indicates that the angular integral was performed over the electric field transmission profile, whereas the I super-index corresponds to a direct integral of the per-ray intensity transmission profile.}}\label{fig:lrehist}
\end{figure*}

We have experimented with the effect of relaxing each of the model requirements for faster calculations. We first kept the full spectral range of the observations, and performed the fits with approximated transmission profile recipes (perpendicular incidence, the \emph{conv} approximation and the full calculation). We then truncated the data (and the model) to only include the central lobe of the HRE and LRE in the fits, for all three transmission profile recipes. Since this limited spectral range does not allow estimating the prefilter parameters, we utilize a much simpler model:
\begin{equation}
I_{\mathrm{obs}} (\lambda) = \left\{ I_{\mathrm{FTS}} (\lambda) \cdot p_g(1.0 + p_{3} \Delta\lambda') \right\} * T_{\mathrm{CRISP2}}(\lambda),\label{eq:hremodel}
\end{equation}
where $p_3$ captures the slope of the observed background continuum, $\Delta \lambda' = \lambda - \lambda_{\mathrm{ref}}$ and $\lambda_{\mathrm{ref}}$ is a reference wavelength, typically the central wavelength of the spectral range under consideration. The main difference appears in the inferred reflectivities of both etalons. For each case, we have assumed that the full profile calculation with the full spectral range is the reference. Figure~\ref{fig:hrehist} depicts 2D density plots of $R_x^{y} - R_{\mathrm{ref}}^{\mathrm{E}}$ vs $R_{\mathrm{ref}}^{\mathrm{E}}$ (where $x$ can refer to the \emph{ray}, \emph{conv} or \emph{full} transmission profile approximations {and $y$ to whether we are using the intensity transmission profile integral $y=I$, or the \emph{complex} electric field transmission profile $y=E$}) and for the cavity separation $C_x^{y} - C_{\mathrm{ref}}^{\mathrm{E}}$ vs $C_{\mathrm{ref}}^{\mathrm{E}}$ . These results show two main effects:
\begin{enumerate}
	\item Using a simpler transmission profile recipe results in a lower fitted reflectivity. The explanation is simple, by ignoring the profile broadening effect of the angular integral (in the perpendicular incidence case), or by ignoring the LRE tilt angle in the \emph{conv} approximation, the fitting routine can only compensate missing broadening by artificially reducing the inferred reflectivity. 
	\item Considering only the central transmission lobe of the profile, while neglecting the contribution of the secondary peaks at $\pm 1\times \mathrm{FSR}$, introduces a larger spread in the inferred reflectivities.The latter is clearly visible when comparing the two columns of Figs.~\ref{fig:hrehist} for each of the transmission profile recipes. 
	\item  {The mean value of the cavity error maps is on average very close to zero when the full spectral range is included. When only the central lobe is considered, the cavity errors have an offset of approximately $0.85$~nm (corresponding to a shift of 67~m\AA\ in wavelength).}
	\item  {The results from calculations using a simpler transmission profile recipe have a larger spread around the mean value. The spread is much larger for the calculations including the central lobe only ($\le 0.02$ to $0.1$~nm vs $\sim 0.2$~nm). }
\end{enumerate}

 {Having a global offset in the derived parameters is manageable. But the spread is more problematic as it captures FOV dependent errors that are harder (perhaps even impossible) to compensate. For example, the \emph{conv} approximation with the full spectral range has little spread in both quantities compared to the reference case, showing a slight offset in the inferred reflectivity that could be calibrated. The full calculation adopting an integral of the intensity transmission profile has even less spread and better captures the asymmetries of the real profile, but the reflectivity offset is larger.}

 {Fig.~\ref{fig:lrehist} shows similar results for the LRE, although in this case we have only considered one spectral range. The LRE is broader and the cavity error determination seems to be less affected by the exact transmission profile recipe that is used in the calculation. The offset is $\sim -0.11$~nm and the spread is very small in all cases ($\le 0.009$~nm). The reflectivity determination is more affected by the usage of simplification that do not capture the tilt of the etalon. Surprisingly, the \emph{conv} calculation predicts a lower reflectivity than the single ray approximation. However, after careful inspection of the quality of the fits (not shown), we conclude that the average value of $\chi^2$ is $\sim 35\%$ larger for the single ray case, reflecting that this formula cannot capture the profile asymmetries induced by the angular integral and the tilt of the LRE.} 
\begin{figure}
	\centering
	\includegraphics[width=\columnwidth]{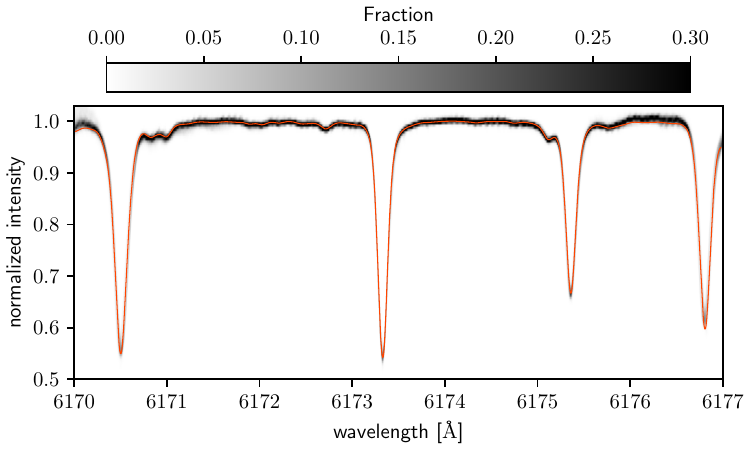}
	\caption{Prefilter-corrected 2D histogram of the observed data. Each wavelength bin has been normalized by the total to better illustrate the spread of the data points. The cavity error map has been compensated in the wavelength array of each $(x,y)$ location in the FOV.  {The red curve corresponds to the FTS atlas (multiplied by the prefilter, convolved with the CRISP2 transmission profile, and divided again by the prefilter curve, also convolved with the CRISP2 transmission profile).}}\label{fig:pcorr}
\end{figure}

\subsection{Fast characterization of the instrumental profile}
 Figure~\ref{fig:hrehist} shows that using a simplified transmission profile calculation can be compensated by (wrongly) adjusting the inferred reflectivity. However, a different question is how different are the resulting CRISP2 profiles from all these calculations. We have computed the transmission profiles using the results of $R_{\mathrm{full}}^{\mathrm{E}}$,  $R_{\mathrm{conv}}^{\mathrm{E}}$ and $R_{\mathrm{full}}^{\mathrm{I}}$ and their corresponding formulas.

\begin{figure}
	\centering
	\includegraphics[width=\columnwidth]{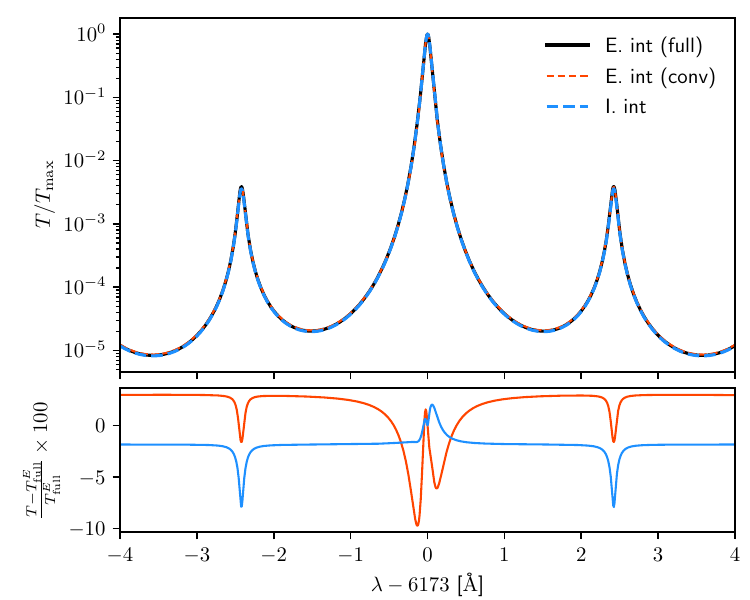}
	\caption{ {CRISP2 transmission profiles calculated from the inferred reflectivities {at the center of the FOV} $(x,y)=(n_x/2,n_y/2)$. The upper panel shows the peak-normalized transmission profiles corresponding to the \emph{full} complex calculation (black), the \emph{conv} complex calculation (red) and the \emph{full} integration of the intensity vector (blue). The bottom panel shows the residues relative to the \emph{full} complex calculation. The residues are normalized by the observed intensity of each curve {per wavelength}.}}\label{fig:profchar}
\end{figure}

 The results are illustrated in Fig.~\ref{fig:profchar}. Our results show that, if the final goal of the fits is to characterize the instrumental profile (for example, for performing inversions), even the faster angular integration of the intensity profile could suffice for this purpose.  {The residues in this figure are defined slightly different than in the rest of the manuscript. Instead of dividing the difference by the peak intensity of each curve, they are normalized by the wavelength-dependent intensity. That way it is much easier to appreciate the differences between profiles across the entire wavelength range, although the reported errors will appear large in regions with very low transmission}. Given this definition, the maximum relative error would be approximately $~7\%$ and generally contained within $\pm 2\%$, however we note that the peak error is found in the secondary transmission lobes where the transmission is very low in comparison to the central lobe. {The two approximate formulas show a systematic offset in the far wings, showing that the decay of the wings is slightly faster for $T_{\mathrm{full}}^I$ and slower for $T_{\mathrm{conv}}^E$}. Using the complex angular integration while ignoring the tilt of the LRE yields larger deviations (similar peak deviations but larger mean deviation).

 These results would suggest that the primary (but not only) source of profile broadening is the contribution from slanted rays in the angular integral, and not phase errors, as relatively small adjustments in the inferred reflectivity can compensate for the latter. Having faster inference algorithms or a faster forward calculation of the instrumental profile can be useful when computational resources are very limited.

\subsection{Flat-fielding and prefilter correction}
The results of the HRE parameterization could be used to generate accurate flats that also contain the prefilter correction all at once. In that case, the flat would be given by Eq.~\ref{eq:hremodel} but without the $I_{\mathrm{fts}}$ factor.
The generation of the gaintables must be done for all wavelength tuning points at once. Such an approach was utilized by \citetads{2013A&A...553A..63S}. Figure~\ref{fig:pcorr} shows the 2D histogram of the HRE dataset divided by the flat-field correction. For this dataset, the expected result after the flat/prefiler correction should be approximately $I_{\mathrm{fts}}\cdot T_{\mathrm{CRISP2}}$. The resulting spectrum has very little spread in the continuum, which is very close to unity over the entire spectral range. In the lines the spread is larger as  {the data in one wavelength bin can be spread over a larger number of intensity bins in the vertical direction. The red curve shows the degraded FTS atlas profile, which is the  reference to evaluate whether the prefilter correction worked fine over the entire wavelength range. With the exception of the blue-most part at $6170$~\AA\ where the prefilter determination is worse, the overall agreement is very good.}

The acquisition of flat-field data over such a large wavelength range is likely not feasible on a daily basis for all spectral lines that are commonly observed together with CRISP2.  {However, adding a few extra continuum points across the prefilter could suffice to estimate the prefilter parameters.}

In comparison, the SSTRED pipeline (\citeads{2015A&A...573A..40D}; \citeads{2021A&A...653A..68L}) estimates a template mean spectrum from the observed flat-field data (instead of using the FTS atlas) and models the FOV-dependent spectra by applying a shift that mimics the cavity error and multiplying by a polynomial that captures the FOV-dependent variations of the prefilter. A global prefilter curve is estimated in a separate step and applied to the reduced data after image reconstruction. This approach is robust in that it does not assume anything about the observed spectrum (it does not need to be similar to the FTS atlas) and it is insensitive to the exact imprint of the tellurics in the FTS spectrum, which will be different than in the observations.

In a near future, we plan to implement the model from \S\ref{sec:simp} in SSTRED.

\subsection{Example dataset}
\begin{figure*}
	\centering
	\includegraphics[width=\textwidth]{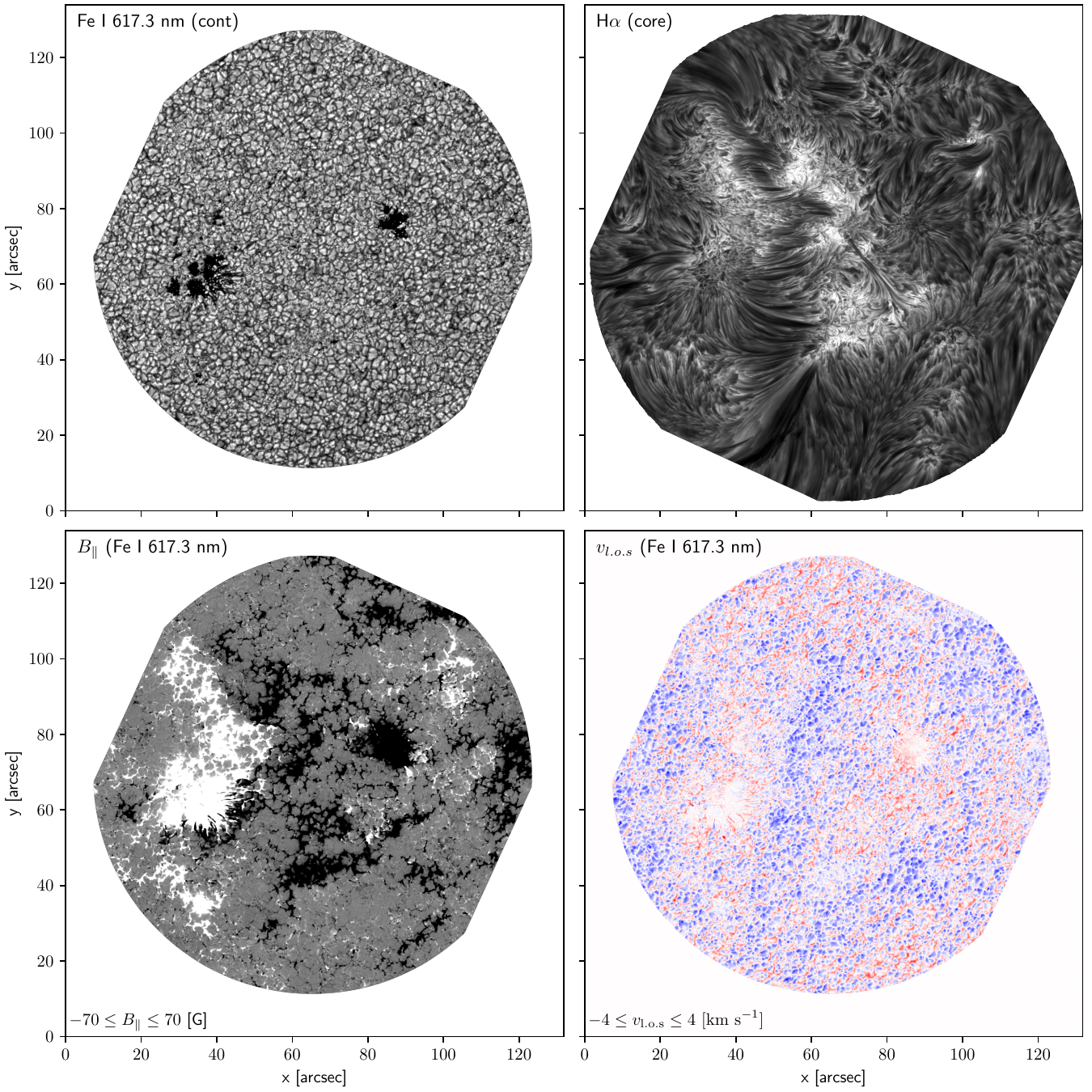}
	\caption{ {CRISP2 dataset acquired in the \ion{Fe}{i}~617.3~nm and H$\alpha$ lines on 2025-10-10 at 09:50:40 UT. The bottom panels show the results of a Milne-Eddington inversion using the 617.3 nm dataset. \emph{Top-left:} continuum image at 617.4~nm. \emph{Top-right}: H$\alpha$ core image. \emph{Bottom-left:} Line-of-sight component of the magnetic field vector, clipped at $\pm 70$~G.\emph{Bottom-right:} Cavity-error-map-compensated line-of-sight velocity, clipped at $\pm 4$~km~s$^{-1}$. \emph{Top-left:} continuum image at 617.4~nm. \emph{Top-right}: H$\alpha$ core image. Data courtesy of A. Brunvoll, R. Nguyen and L. Rouppe van der Voort (UiO).}}\label{fig:halpha}
\end{figure*}

 {Figure~\ref{fig:halpha} depicts two quasi-monochromatic images acquired with the CRISP2 instrument in the 617.3~nm continuum and in the core of the H$\alpha$ line. Additionally, we have performed a regularized (in space and time) Milne-Eddington inversion (\citeads{2019A&A...631A.153D}; \citeads{2024A&A...685A..85D}) of the \ion{Fe}{i}~617.3~nm dataset. We display the derived line-of-sight components of the magnetic and velocity fields (cavity-error corrected). The data have been processed with the SSTRED pipeline (\citeads{2021A&A...653A..68L}; \citeads{2015A&A...573A..40D}) and the multi-object-multi-frame-blind-deconvolution image reconstruction technique (MOMFBD, \citeads{2005SoPh..228..191V}; \citeads{2002SPIE.4792..146L}). These data and inversion results show no systematic errors or artifacts that could indicate poor performance of the instrument.}

 {Although the data shows an unprecedentedly large FOV for these type of instrument and spatial resolution, at the moment the FOV is limited by the size of the modulator and the 617.3~nm prefilter. Both will be replaced during 2026 with larger versions that will allow exploiting the entire FOV that is physically available.}

\section{Conclusions}
We have proposed a method (along with several simplifications) that allows for a full characterization of dual-etalon FPI systems, including the parameters of order-selecting prefilters, from observational data. Therefore, it can be utilized at any wavelength for which an order-selecting prefilter is available without changing the optical setup. Compared to laser-based measurements, where the laser is placed on the optical table, our approach has the advantage of including the broadening effect of the slowly-converging telecentric beam, which cannot be captured by such laser measurements. 

We have performed a full characterization of the new CRISP2 dual-etalon FPI parameters at 617.3~nm. Our results show that the FOV-dependent cavity separation errors are small, with RMS values $\sim2.0$~nm for both etalons. The etalon reflectivities have RMS variations of $0.45\%$ and $0.33\%$ for the HRE and LRE at $617.3$~nm. 
At other wavelengths, the overall field-dependence should behave similarly, although the average reflectivity will obviously change.

We have assessed the effect of inferring the FPI parameters with simplified transmission profile approximations and with a reduced spectral coverage. The former systematically leads to lower inferred reflectivities whereas the latter also introduces a large dispersion in the inferred reflectivities in addition to the offset. Our results suggest that the \emph{conv} approximation can be used in the determination of HRE reflectivities at a minimal accuracy cost. Our recommendation is to use, at the very least, the \emph{conv} approximation in the HRE parameter determination, which only requires a small set of incident angles and yields similar results to the \emph{full} calculation. For the LRE, the model has fewer parameters and the restricted spectral range allows for fast estimation using the \emph{full} approach.

 {If the only goal is to characterize the FOV-dependent transmission profile of the system, without further interpretation of the derived etalon reflectivities, our results suggest that the faster (approximated) angular integration of the intensity transmission profile yields very similar results to the full complex integration of the electric field transmission profile. In this case, the reflectivity is modified in order to compensate the inaccuracies in the profile calculation, but the resulting transmission profiles are virtually identical in shape and area (but not in peak transmission).}

We show that an estimate of the prefilter curve across the FOV is generally required to characterize the reflectivities of the HRE and LRE systems. However, if the prefilter is sufficiently narrow in comparison with the FSR of the FPI, the secondary transmission lobes will be damped when observing at the center of the prefilter. Our results suggest that such cases can be reasonably modeled without the inclusion of the transmission peaks at $\pm 1\times$FSR. Regular flat-field data could be used in those cases.

Modern inversion codes allow specifying a FOV-dependent  {spectral transmission profile}, for example: NICOLE (\citeads{2015A&A...577A...7S}), SNAPI (\citeads{2018A&A...617A..24M}),  STiC (\citeads{2019A&A...623A..74D}), FIRTEZ-dz (\citeads{2019A&A...629A..24P}). This feature is very important due to the FOV-dependent variations of the spectral  {transmission profile} (wavelength shift and broadening variations) that are imprinted in FPI observations.

All the codes used in this study have been made publicly available\footnote{\url{https://github.com/jaimedelacruz/pyFPI}} with documented examples. We hope they become useful to the solar community in the characterization of future telecentric FPI instruments at the Daniel K. Inouye Solar Telescope \citepads{2020SoPh..295..172R} and at the European Solar Telescope \citepads{2022A&A...666A..21Q}.

% %%%%%%%%%%%%%%%%%%%%%%%%%%%%%%%%%%%%%%%%%%%%%%%%%%%%%%%%%%%%%%

\begin{acknowledgements}
% SST
The Institute for Solar Physics is supported by a grant for research infrastructures of national importance from the Swedish Research Council (registration number 2021-00169). The Swedish 1-m Solar Telescope is operated on the island of La Palma by the Institute for Solar Physics of Stockholm University in the Spanish Observatorio del Roque de los Muchachos of the Instituto de Astrof\'isica de Canarias.
This project has been funded by the European Union through the European Research Council (ERC) under the Horizon Europe program (MAGHEAT, grant agreement 101088184). The European Solar Telescope project is supported by a grant for research infrastructures from the Swedish Research Council (registration number 2023-00169). Code debugging was possible thanks to resources provided by the National Academic Infrastructure for Supercomputing in Sweden (project NAISS 2025-1-9) at the PDC Centre for High Performance Computing (PDC-HPC) at the Royal Institute of Technology in Stockholm. We are very grateful to L. Rouppe van der Voort for providing the dataset shown in Fig.~\ref{fig:halpha}.
 \end{acknowledgements}
 
%%%%%%%%%%%%%
% %%%%%%%%%%%%%%%%%%%%%%%%%%%%%%%%%%%%%%%%%%%%%%%%%%%%%%%%%%%%%%

\bibliographystyle{aa}
\bibliography{references}

\onecolumn

\begin{appendix}
\section{Assumption of a constant transmission profile in the models}\label{ap:conv}
\begin{figure}[!ht]
	\centering
	\includegraphics[width=\columnwidth]{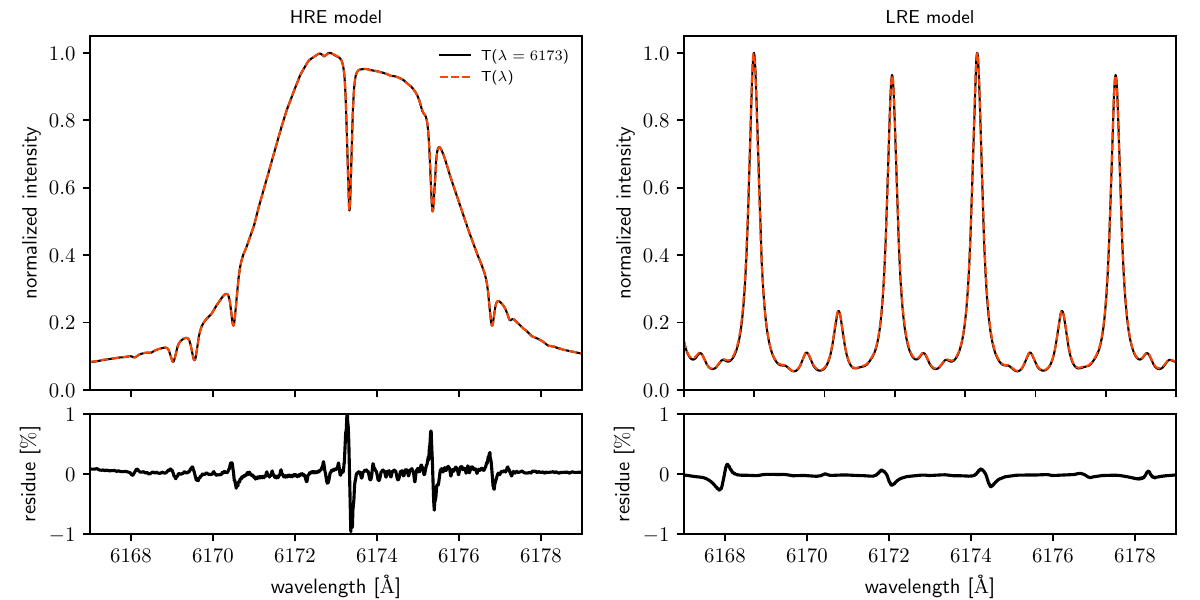}
	\caption{Simulated data assuming a constant transmission profile calculation (black curve, calculated at $\lambda=6173$~\AA) and a wavelength varying transmission profile (red curve). The latter was recalculated for each wavelength point with updated reflectivity and wavelength values. \emph{Left:} Simulated HRE dataset using the FTS atlas and a symmetric prefilter curve centered at $6173$~\AA\ and a FWHM of  $4.8$. \emph{Right:} Simulated LRE dataset using the same prefilter parameters as in the HRE calculation. The residues are normalized by the peak observed intensity of each curve.}\label{fig:convwave}
\end{figure}

 {The transmission profile of an etalon is wavelength dependent. This dependency is contained in the phase term and in the wavelength variation of the etalon reflectivity.
To speed-up calculations, we have assumed that the CRISP2 transmission profile can be assumed to be constant within the wavelength range covered by the prefilter. We have assessed the validity of this assumption in both HRE and LRE models. Figure~\ref{fig:convwave} illustrates the results of performing the convolutions in both models with a constant transmission profile as a function of wavelength (black curve) and a wavelength-dependent transmission profile (red curve). }

Within the range of the prefilter in the HRE model, errors are  $\le \pm 1\%$. The approximations adopted in the LRE model yield a prediction with a maximum discrepancy of $\pm0.2\%$ in comparison with that including a wavelength-varying instrumental profile. We argue that the assumption of a wavelength-invariant transmission profile within the wavelength ranges considered in our study is a good approximation that allows for a much quicker processing of very large FOVs.

\section{Analytical derivatives of the FPI transmission profile}\label{ap:der}
For any given incidence angle $\theta$, the derivative of the transmission profile of an etalon (relative to the cavity separation and reflectivity, see Eq.~\ref{eq:tr1}) can be trivially estimated:
\begin{flalign}
\frac{\partial T(C,R,\lambda)}{\partial R} &= \frac{T}{1-R} + \frac{1}{1-R} \left [ \frac{-\cos(\psi/2) + i\sin(\psi/2)}{1+F\sin^2(\psi/2)} - T\frac{\sin^2(\psi/2)}{1+F\sin^2(\psi/2)} \frac{4(1+R)}{(1-R)^3}\right],\label{eq:dR}\\
\frac{\partial T(C,R,\lambda)}{\partial C} &= \left [\frac{1}{1-R}\frac{(R-1)\sin(\psi/2) + i(1+R)\cos(\psi/2)}{1+F\sin^2(\psi/2)} - T\frac{2F\sin(\psi/2)\cos(\psi/2)}{1+F\sin^2(\psi/2)}\right ] \frac{2\pi n \cos(\theta) }{\lambda},
\end{flalign}
where $T$ is the transmission profile (also in the right-hand-side).

\begin{figure}[!ht]
	\centering
	\includegraphics[width=\columnwidth]{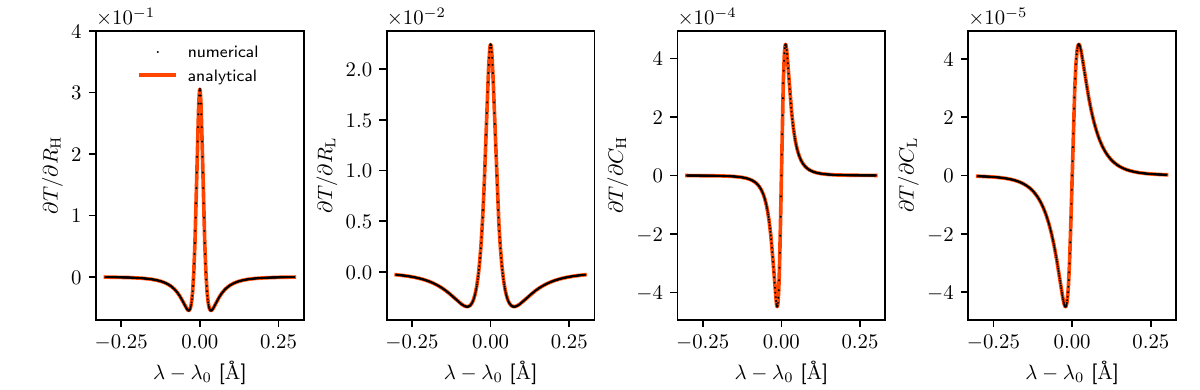}
	\caption{Derivatives of the area-normalized transmission profile relative to the reflectivity ($R$) and cavity separation ($C$) of each of the etalons. The analytical calculation is depicted in red and the numerical ones are indicated with black dots.}\label{fig:tder}
\end{figure}

We note though that the resulting FPI profile is not area normalized. The $\sin(\psi/2)$ and $\cos(\psi/2)$ terms in Eq.~\ref{eq:tr1} generate an infinite succession of transmission peaks that has no general analytical integral. But assuming that we are operating on a regular (discrete) wavelength grid within a finite wavelength domain, the area of \emph{that} transmission profile can be estimated as $A_{tr} =  \sum_i T(C,R,\lambda_i)$ (dropping the $\delta\lambda$ factor as it vanishes in the convolution). The normalized (discrete) profile is given by:
\begin{equation}
T_n = \frac{T(C,R)}{A_{tr}}.
\end{equation}
The only thing we need to know is the derivative of the profile area relative to the cavity separation and the etalon reflectivity. For very small perturbations, changing the cavity separations only produces a displacement of the transmission profile and the area remains the same. The reflectivity derivative is already given in Eq.~\ref{eq:dR} and it is wavelength dependent. By defining $B_{tr}=\sum_i \partial T(C,R,\lambda_i)/\partial R$, we can easily calculate the derivative of the area normalized profile using the chain rule as:
\begin{equation}
\frac{\partial T_n(C,R)}{\partial R} = \frac{1}{A_{tr}}\left(\frac{\partial T(C,R)}{\partial R} -\frac{ B_{tr} }{A_{tr}}T(C,R)\right).
\end{equation}
Figure~\ref{fig:tder} depicts a comparison between analytical and numerical derivatives of the area-normalized transmission profile at 630.2~nm using the nominal reflectivities and cavity separations of the etalons. 

\section{Analytical derivatives of the prefilter transmission profile}\label{ap:prefvar}
\begin{figure}
	\centering
	\includegraphics[width=\columnwidth]{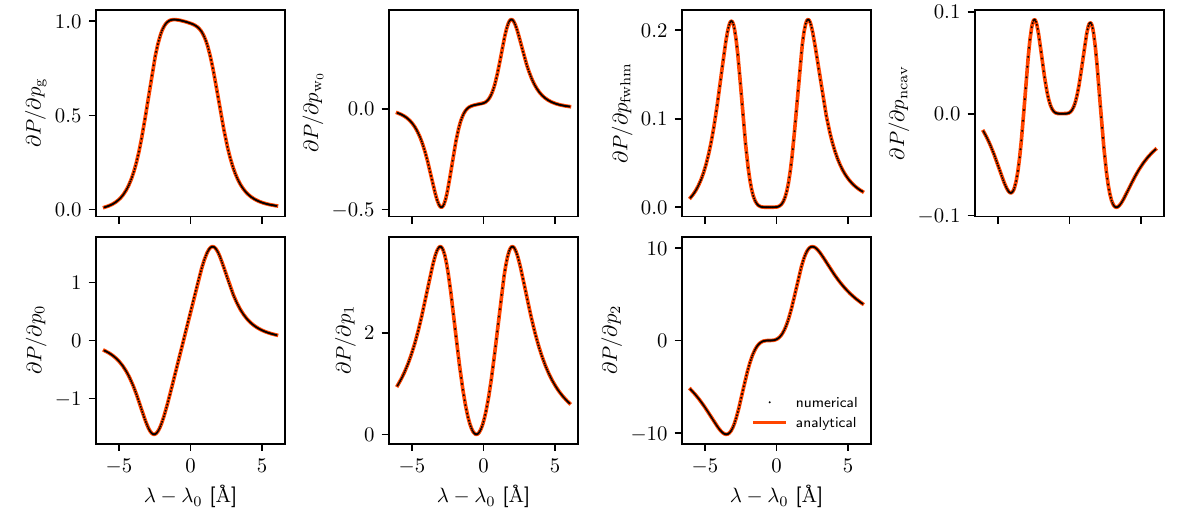}
	\caption{Derivatives of the prefilter curve with respect to all seven parameters considered in Eq.~\ref{eq:pref}. The red curve illustrates the analytical derivatives and the black dots the finite-differences ones. These derivatives include the effect of the apodization window.}\label{fig:pder}
\end{figure}

The derivatives of the prefilter parameters can be easily derived in analytical form, which we include here for completeness. For a given $\lambda$ and $\Delta\lambda=\lambda-p_{w_0}$:
\begin{flalign} 
\frac{\partial P}{\partial p_g} &= \frac{1}{1 + q^{2p_{\mathrm{ncav}}}}  \left(1 + w_{\mathrm{apod}}(p_0 \Delta\lambda+ p_1 \Delta\lambda^2 + p_2 \Delta\lambda^3) \right), \\
\frac{\partial P}{\partial p_{w_0}} &=  2\frac{p_g}{\left( 1 + q^{2p_{\mathrm{ncav}}}\right)^2}\frac{q^{2p_{\mathrm{ncav}}-1}\cdot \sign(q)}
{p_{\mathrm{fwhm}}}  \left(1 + w_{\mathrm{apod}}(p_0 \Delta\lambda+ p_1 \Delta\lambda^2 + p_2 \Delta\lambda^3) \right)- \frac{p_gw_{\mathrm{apod}}}{1 + q^{2p_{\mathrm{ncav}}}} \left(p_0 + 2p_1\Delta\lambda +3p_2\Delta\lambda^2\right ) ,\\
\frac{\partial P}{\partial p_{\mathrm{fwhm}}} &=  \frac{p_g}{\left( 1 + q^{2p_{\mathrm{ncav}}}\right)^2} \frac{q^{2p_{\mathrm{ncav}}-1}\cdot |q|}
{p_{\mathrm{fwhm}}}  \left(1 + w_{\mathrm{apod}}(p_0 \Delta\lambda+ p_1 \Delta\lambda^2 + p_2 \Delta\lambda^3 )\right) ,\\
\frac{\partial P}{\partial p_0} &=\frac{p_g}{1 + q^{2p_{\mathrm{ncav}}}}  w_{\mathrm{apod}}\Delta\lambda,\\
\frac{\partial P}{\partial p_1} &=\frac{\partial P}{\partial p_0} w_{\mathrm{apod}}\Delta\lambda, \\
\frac{\partial P}{\partial p_2} &=\frac{\partial P}{\partial p_1} w_{\mathrm{apod}}\Delta\lambda, \\
\end{flalign}
where $q = 2(\lambda - p_{w_0})/p_{\mathrm{fwhm}}$, and $\Delta \lambda = \lambda-p_{w_0}$. Fig.~\ref{fig:pder} illustrates the derivatives of the prefilter curve relative to all parameters. 

Our definition of the apodization window is analytical, differentiable and it has a dependence on $p_{w_0}$ and $p_{\mathrm{fwhm}}$. For a given $\Delta\lambda = \lambda-p_{w_0}$, the derivatives of the apodization window are:
\begin{flalign}
\frac{\partial w_{\mathrm{apod}}}{\partial p_{w_0}} &=  -\frac{\pi}{2a_{\mathrm{scl}}p_{\mathrm{fwhm}}}\left((1-\tanh^2(\Delta\lambda_1)\cdot(1-\tanh(\Delta\lambda_2)) + (1+\tanh(\Delta\lambda_1))\cdot (\tanh^2(\Delta\lambda_2)-1)\right),\\
\frac{\partial w_{\mathrm{apod}}}{\partial p_{\mathrm{fwhm}}} &= -\frac{\pi}{2a_{\mathrm{scl}}}\frac{(\Delta\lambda)\cdot\sign(\Delta\lambda)}{p_{\mathrm{fwhm}}^2}\left((1-\tanh^2(\Delta\lambda_1)\cdot(1-\tanh(\Delta\lambda_2)) - (1+\tanh(\Delta\lambda_1))\cdot (\tanh^2(\Delta\lambda_2)-1)\right).
\end{flalign}
Therefore, this dependence must be propagated to the derivatives of the prefilter curve relative to $p_{w_0}$ and $p_{\mathrm{fwhm}}$ by adding the corresponding terms:
\begin{flalign}
\left.\frac{\partial P}{\partial p_{w_0}}\right\rvert_{\mathrm{full}} &=\frac{\partial P}{\partial p_{\mathrm{w_0}}} + \frac{p_g}{1 + q^{2p_{\mathrm{ncav}}}} \cdot(p_0 \Delta\lambda+ p_1 \Delta\lambda^2 + p_2 \Delta\lambda^3)\frac{\partial w_{\mathrm{apod}}}{\partial p_{w_0}}, \\
\left.\frac{\partial P}{\partial p_{whm}}\right\rvert_{\mathrm{full}} &=\frac{\partial P}{\partial p_{\mathrm{fwhm}}}+\frac{p_g}{1 + q^{2p_{\mathrm{ncav}}}} \cdot(p_0 \Delta\lambda+ p_1 \Delta\lambda^2 + p_2 \Delta\lambda^3)\frac{\partial w_{\mathrm{apod}}}{\partial p_{\mathrm{fwhm}}}.
\end{flalign}

\end{appendix}

%\begin{appendix}
%\input{appendix.tex}
%\end{appendix}

\end{document}